\documentclass[%
 reprint,
 amsmath,amssymb,
 aps,
]{revtex4-1}

\usepackage{graphicx}% Include figure files
\usepackage{dcolumn}% Align table columns on decimal point
\usepackage{bm}% bold math
\newcommand{\q}{\quad}
\newcommand{\qq}{\qquad}
\newcommand{\bea}{\begin{eqnarray}}
\newcommand{\eea}{\end{eqnarray}}
\newcommand{\ket}[1]{\left|{#1}\right\rangle}
\newcommand{\bra}[1]{\left\langle{#1}\right|}
\newcommand{\aver}[1]{\left\langle{#1}\right\rangle}
\newcommand{\expect}[3]{\left\langle{#1}\right|{#2}\left|{#3}\right\rangle}

\newcommand{\inner}[2]{\left\langle{#1}|{#2}\right\rangle}

\usepackage{hyperref}
\usepackage{caption} 
\usepackage{booktabs,siunitx}

\usepackage{enumerate}
\usepackage{float}

\begin{document}

%\preprint{APS/123-QED}

\title{Nonclassical effects in optomechanics: Dynamics and collapse of entanglement}% Force line breaks with \\

\author{Pradip Laha}
 \altaffiliation[]{pradip@physics.iitm.ac.in}%Lines break automatically or can be forced with \\
\author{S Lakshmibala}%
 \email{slbala@physics.iitm.ac.in}
\affiliation{Department of Physics, IIT Madras, Chennai 600036, India\\}%
\author{V Balakrishnan}%
 \email{vbalki@physics.iitm.ac.in}
\affiliation{Department of Physics, IIT Madras, Chennai 600036, India\\}%


%\affiliation{
%\affiliation{
%\affiliation{%


\date{\today}% It is always \today, today,

\begin{abstract}
We have investigated a wide range of nonclassical behavior exhibited by a tripartite cavity optomechanical system comprising a two-level atom placed inside a Fabry-P\'{e}rot type optical cavity with a vibrating mirror attached to one end. We have shown that the  atom's  subsystem von Neumann entropy collapses to its maximum allowed value over a significant time interval during dynamical evolution. This feature is  sensitive to the nature of the initial state, the  specific form of intensity-dependent tripartite coupling,  and system parameters. The extent of nonclassicality of the field is assessed through the Mandel Q parameter and Wigner function. Both entropic and quadrature  squeezing properties of the  field are quantified directly from  optical tomograms, thereby  avoiding tedious  state reconstruction procedures. 
\end{abstract}

\pacs{Valid PACS appear here}% PACS, the Physics and Astronomy
\maketitle

%\tableofcontents

%%%%%%%%%%%%%%%%%%%%%%%%%%%%%%%%%%%%%%%%%%%%%%%%%%%%%%%%%%%%%%%
\section{Introduction}
\label{Introduction}
In recent years, the dynamical behavior  of optomechanical systems   has attracted  considerable attention (see, for instance, \cite{aspelmeyer,bowen}). In cavity optomechanics, the basic model involves the interaction between the optical field contained  in a cavity and  a mechanical oscillator whose motion is due to the  radiation pressure.  Controlling the dynamics of   a quantum oscillator in this manner has found interesting applications in the   detection of gravitational waves \cite{abramovici,braginsky}, high precision measurements of masses and the weak force \cite{vitali2,geraci,lamoreaux},  the processing of quantum information \cite{stannigel}, cooling mechanical resonators very close to their quantum ground states \cite{barzanjeh,wilson_rae,genes1,li},  
and examining the transition between classical and quantum behavior of a mechanical system \cite{schwab,marshall}.  

 In contrast, levitated optomechanics,  where the cavity is dispensed with and a nano-particle is subjected  to radiation pressure, provides an excellent platform for minimising dissipation effects. Interesting results from a series of experiments on such a system have been reported in the literature, including reconstruction of the Wigner function of the particle \cite{toros1} and  tracking of the rotational and translational dynamics of an anisotropic particle \cite{toros2}.

Theoretical investigations on the entanglement dynamics exhibited in cavity optomechanics have been carried out on a variety 
of these  systems.
In an atomic ensemble surrounded by a high-finesse optical cavity with an attached vibrating mirror,  both bipartite and tripartite entanglements have been investigated in experimentally accessible parameter regimes \cite{genes2}. A modified version comprises a single two-level atom placed inside the  cavity to which  a vibrating mirror is  attached at one end.  In this case tripartite entangled states have been examined. In Ref. \cite{liu_ijtp}, 
for instance,   an initial factored product state  of a single photon, the first excited state 
of the oscillator and the excited state of the atom 
has been  shown to transform to a Greenberger-Horne-Zeilinger (GHZ)-like entangled state at a subsequent instant. The occurrence  of sudden entanglement birth and death during dynamical evolution 
 has been noted, and the effect of dissipation has been studied. The manner in which the degree of atomic coherence, the coupling strengths and system parameters can be exploited to control entanglement in the absence of dissipation has been reported in \cite{liao_laser}.  An extension of this model  has been  studied in \cite{hassani}, to examine the interaction of a $V$-type atom with a two-mode quantized field. The possibility of strong coupling between the quantized motion of a mechanical oscillator and a multi-level trapped atom, both initially close to their respective ground states in a cavity optomechanical set-up, has been considered in \cite{hammerer}. The role of dissipation in 
a  strongly-coupled field-atom system has been examined  in \cite{wallquist}. Almost all of these investigations have been carried out for unentangled initial states with the field in a  specific  photon number state.

 A new dimension to these investigations arises with the incorporation of an intensity-dependent coupling (IDC).  The effect of a nonlinear tripartite field-atom-oscillator coupling term of the form $(1-\tfrac{1}{2} \kappa \,b^{\dagger} b)$ (where $\kappa$ is the tunable  intensity parameter and $b$, $b^{\dagger}$ are the oscillator ladder operators) on the system dynamics 
 has been  analysed  in \cite{barzanjeh}. This particular form of  coupling is attributed to the spatial field-mode structure at the position of the two-level atom inside the cavity.  The role played by different forms of IDCs in more general settings of field-atom interactions has  also been examined. These include couplings of the form  $(a^{\dagger}a)^{1/2}$ \cite{buck},  $(a^{\dagger}a)^{-1/2}$ \cite{sudarshan}  and $(1 + \kappa\, a^{\dagger}a)^{1/2}$ \cite{laha1} where the parameter $\kappa$ takes values in the range $[0, 1]$, and $a$, $a^{\dagger}$ are field ladder operators. This last form of intensity dependence is interesting from  a group-theoretic point of view. There is an underlying algebraic structure for the field operators associated with this particular functional form of the coupling. Two limiting cases are of particular interest: the case  $\kappa = 0$ which reduces to the Heisenberg-Weyl algebra for the field operators, and the case  $\kappa = 1$ which leads to the SU(1, 1) algebra for nonlinear combinations of these operators \cite{siva2}; intermediate values of $\kappa$ correspond to a deformed SU(1, 1) operator algebra.

 In \cite{laha1}, this last form of IDC between a $V$ or $\Lambda$ atom and two radiation fields (that mediate allowed  transitions between the two pairs of atomic states)  has been shown to  lead to the occurrence, during 
 dynamical evolution,  of  a bifurcation cascade that  is very sensitive to the precise value of $\kappa$.  More significantly,  it enables collapse of  a specific bipartite entanglement   to a non-zero value over a significant time interval during the system's temporal evolution.  In view of the fact that  it is possible to minimise dissipation effects in optomechanics, it would be 
 very  useful to identify the occurrences of such collapses of entanglement to constant non-zero values in optomechanical systems.  Accordingly, we undertake in this paper a detailed investigation of the role played by various forms of intensity-dependent couplings in a generic model of cavity optomechanics. 
 
A  novel feature we find  is the following: for specific experimentally accessible parameter values,   the effective bipartite entanglement collapses to its maximum  possible value over a substantial time interval. The degree of entanglement  as quantified by the subsystem von Neumann entropy (SVNE) S$_{\text{i}} = -\textrm{Tr}\, [\rho_{i} \ln \rho_{i}]$ (where $i$ is the subsystem label) corroborates  this collapse property.  This feature also manifests 
 itself in  the dynamics of the mean photon number $\aver{N}$, the corresponding variance,  and the  Mandel parameter 
 $Q = \aver{N}^{-1}\aver{(\Delta N)^{2}} - 1$.  ($Q < 0$ 
 signifies  sub-Poissonian statistics or nonclassicallity of the field.)  

  We have  also analysed, quantitatively, the squeezing properties  of the cavity field. This has been carried out {\it directly} from the optical tomogram of the field, circumventing  explicit state reconstruction, and thereby  minimising  statistical errors  that 
  are inevitable during reconstruction.  Optical tomograms are essentially histograms obtained  from homodyne measurements of a quorum of observables \cite{manko1,manko2}. In principle, the optical tomogram contains all the information about the system, and is an alternative representation of the quantum state \cite{manko_ibort}.
The advantage of this approach is borne out by several investigations in recent years involving  optical tomography. 
These include: the  identification of squeezed light and other nonclassical states of light \cite{smithey,schiller}; obtaining  qualitative signatures of revivals and fractional revivals of the initial state of a system with a nonlinear Hamiltonian \cite{rohithpra,saumitran}; and  determining  whether a bipartite state is entangled at the output port of a quantum beamsplitter  for a specific choice of input states \cite{rohithjosab,laha2}, directly from the relevant tomogram.  We examine both  the quadrature and tomographic entropic squeezing properties of the field states and report results on the crucial role played by various forms of IDC on the degree of squeezing.

The plan of the rest of this paper is as follows. In Section 
\ref{model} we describe the physical model studied. In Section \ref{result} we investigate, in detail, the dynamics of  entanglement as 
exhibited in the SVNE of various subsystems, the effect of different forms of IDC on the SVNE, the nonclassicality of the field as displayed in the Mandel Q parameter, and the field Wigner functions at appropriate instants of time. In Section \ref{tomo_section}, we examine the dynamics of both quadrature and entropic squeezing properties of the field from optical tomograms. 
The final section is devoted to a few  brief concluding remarks.  
Some  relevant technical details of the calculations are 
outlined in a set of appendices. 
In  Appendix \ref{appendix_eff_ham}, the procedure for 
obtaining an effective Hamiltonian for the system under consideration is sketched. 
Expressions for the state vector corresponding to the total system and the subsystem density matrices are derived in 
 Appendix \ref{appendix_state_vector}. 
The key  steps in the derivation of the Wigner density of the field 
are indicated in Appendix \ref{appendix_wigner_function}. 

%%%%%%%%%%%%%%%%%%%%%%%%%%%%%%%%%%%%%%%%%%%%%%%%%%%%%%%%%%%%%%%%%%%
 \section{The tripartite cavity optomechanical model}
 \label{model}
\begin{figure}[h]
\centering
\includegraphics[height=3.8cm, width=7cm]{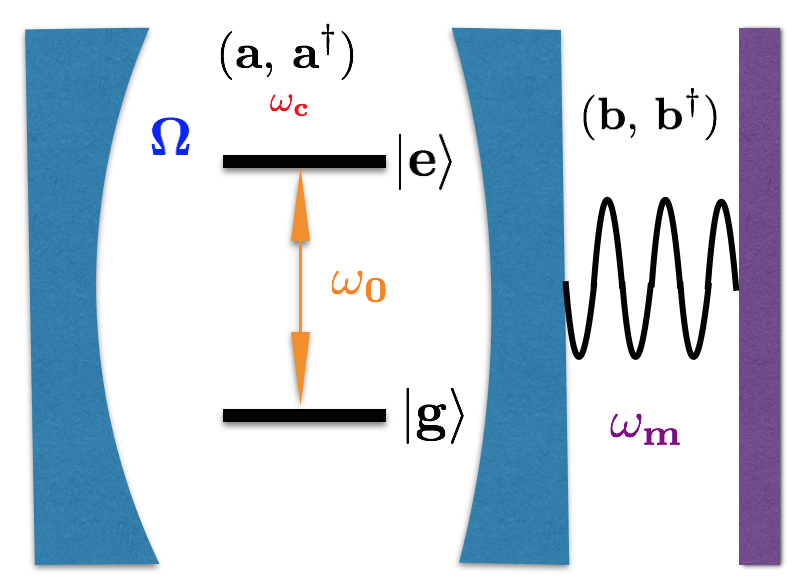}
\caption{Schematic diagram of a cavity optomechanical system.}
\label{cavity_model}
\end{figure}

  The system comprises a two-level atom placed inside a Fabry-P\'{e}rot type optical cavity with a vibrating mirror attached to one end  (figure \ref{cavity_model}). The  mirror is modelled as a quantum harmonic oscillator. The model Hamiltonian  (setting $\hbar$ = 1) is given by
\begin{align}
  H &= \omega\, a^{\dagger} a + \omega_{m}\, b^{\dagger} b + \tfrac{1}{2}\omega_{0} \sigma_{z} - G \,  a^{\dagger} a (b + b^{\dagger}) \nonumber \\
                                           &\qq\qq\qq\qq + \Omega\, [a\, f(N) \, \sigma_{+} +  f(N) \,a^{\dagger}\,\sigma_{-}].
 \label{eqn:parent_hamiltonian}
\end{align}
Here $a^{\dagger},\,a $ and $b^{\dagger},\,b$ are,  respectively, 
the  creation and annihilation operators of the cavity mode with frequency $\omega$ and the mirror-oscillator unit with frequency $\omega_{m}$.  $G = (2\,m\,\omega_{m}/\hbar )^{-1/2}\, (\omega/L)$ is the optomechanical coupling coefficient, where  $L$ and $m$ are the length of the cavity and the mass of the  mirror. $\sigma_{z} = \ket{e}\bra{e} - \ket{g}\bra{g}$, \,$\sigma_{+} = \ket{e}\bra{g}$ and $\sigma_{-} = \ket{g}\bra{e}$, where 
  $\ket{e}$ and $\ket{g}$ are, respectively,  the excited and ground states of the atom. $\omega_{0}$ is the atomic transition frequency and $\Omega$ is the field-atom coupling constant. In our analysis, 
  we have used the resonance condition $\omega = \omega_{0} + \omega_{m}$.  The real-valued function 
  $f(N)$ (where $N = a^{\dagger}a$) incorporates field-atom intensity-dependence.

 As shown in  
 Appendix \ref{appendix_eff_ham}, 
 an  effective Hamiltonian $H_{\textrm{eff}}$ for this system can be obtained from $H$ in the limit $\omega_{m} \gg G, \,\Omega$. It is given by 
 \begin{align}
 H_{\textrm{eff}} = & \frac{G \Omega}{\omega_{m}} \Big[f(N) a^{\dagger}\, b\, \sigma_{-} + a f(N)b^{\dagger} \, \sigma_{+}\Big]  \nonumber \\
                             & - \frac{\Omega^{2}}{\omega_{m}} \Big[a^{\dagger} a\, \sigma_{z} - \sigma_{+}\sigma_{-}\Big] -  \frac{G^{2}}{\omega_{m}} \,(a^{\dagger} a)^{2}.
\label{eqn:eff_hamiltonian}
\end{align}
Note the emergence in $H_{\textrm{eff}}$  of (a) the intensity-dependent tripartite interaction between the atom, field and mirror (the terms proportional to $f(N)$ on the right-hand side), and (b) the  Kerr nonlinearity in $H_{\textrm{eff}}$ (the last term on the right-hand side),   although neither of  these features is explicit in  $H$. We start with an unentangled initial state of the system that is a direct product of  the following states: (i)  the field in a general superposition $\sum_{n=0}^{\infty} l_{n}\ket{n}$ of photon number states $\ket{n}$ (in contrast to \cite{liao_laser}); (ii) the mirror in the oscillator ground state $\ket{0}$;  and  (iii) the atom in an arbitrary superposition $(\cos\phi \ket{e} + \sin\phi \ket{g})$. Thus 
\begin{equation}
 \ket{\psi(0)} = \sum_{n=0}^{\infty} l_{n} (\cos\phi \ket{n; 0; e} + \sin\phi \ket{n; 0; g}),
\label{eqn:init_state}
\end{equation}
in an obvious  notation. Solving the Schr\"{o}dinger equation in the interaction picture, the state of the system at any instant of time is given by
\begin{align}
\ket{\psi(t)} = \sum_{n=0}^{\infty}&  l_{n} A_{n}(t) \ket{n; 0; e} + \sum_{n=0}^{\infty}  l_{n} B_{n}(t) \ket{n; 0; g} \nonumber \\
                                                             &\qq\qq\q+ \sum_{n=1}^{\infty} l_{n}C_{n}(t) \ket{n-1; 1; e},
\label{eqn:final_state}
\end{align}
where explicit expressions for the coefficients $A_{n}(t), B_{n}(t), C_{n}(t)$ and for the subsystem density operators are given in Appendix \ref{appendix_state_vector}.
 
%%%%%%%%%%%%%%%%%%%%%%%%%%%%%%%%%%%%%%%%%%%%%%%%%%%%%%%%%%%%%%%%%%%%%%%%%%
\section{Entanglement dynamics}
 \label{result}
Let us now apply the foregoing to the case when the  initial state of the field is a coherent state (CS) $\ket{\alpha}$ ($\alpha\in\mathbb{C}$), so that $l_{n} = \exp\,(- |\alpha|^{2}/2) \, (n!)^{-1/2}\, \alpha^{n}$. As we shall see, the dynamics is very sensitive to the specific form of the intensity dependence $f(N)$ of the tripartite coupling.  Interesting features are exhibited by the entanglement (as characterized by the SVNE), squeezing properties,  and the Wigner functions at specific  instants of time.
  
In the investigations that follow, we choose experimentally realizable values of the relevant parameters. The typical cavity length $L$ is of the order of  
$\mu$m.  Experiments  have been carried out \cite{hood}   
with cesium atoms passing through a cavity of length $L = 10\,\mu$m, with atomic transition (6S$_{1/2}$, $F=4$, $m_{F} = 4\longrightarrow$ 6P$_{3/2}$, $F=5$, $m_{F} = 5$) and $\omega \sim 10^{14}$ Hz. 
Further, in such a set-up the mass $m$ of the oscillator is of the order of  $10^{-17}$ kg, and the  corresponding  oscillator frequency $\omega_{m}$ is of the order of $10^{9}$ Hz \cite{cleland}. From this it follows that the value of the optomechanical coupling coefficients $G$ is $10^{6}$ Hz, which is much smaller than the value of $\omega_{m}$. The resonance condition  $\omega = \omega_{0} + \omega_{m}$  is satisfied.
Further, the coupling $\Omega$ between the atom and the cavity depends on the atomic position $r$ through the relation $\Omega = \Omega_{0}\, \exp(-r^{2}/w_{0}^{2})$, with the maximum value of the vacuum-Rabi frequency $\Omega_{0} = 2\pi \times 120$ MHz and the waist of the cavity mode $w_{0}\approx 15 \mu$m \cite{hood}. Hence, by adjusting $r$, 
the value of 
$\Omega$ can be set to be close to that  of $G$. We examine two possibilities here, namely, \, (a) \,$G = \Omega$ and \,(b)\, $G = \sqrt{2}\,\Omega$. The latter choice is considered in order to examine whether qualitative features of the system  dynamics are sensitive to changes in the ratio $G/\Omega$, because it has been reported earlier \cite{liao_laser} that the dynamics of the subsystem entropy is qualitatively different for these two values of the ratio concerned. We have examined in the following sections the manner in which both  entropic and quadrature squeezing depend on  the value of $G/\Omega$ during dynamical evolution. Since $G$ and 
$\Omega$ are comparable in their numerical values, the effective frequency (\eqref{eqn:eff_hamiltonian}) is  given by $(G^{2}/\omega_{m})$. 
It is therefore natural to examine the dynamics in terms of 
the dimensionless time variable $\tau = (G^{2}/\omega_{m})t$. 

\subsection{Intensity-independent tripartite coupling}
\label{intn_indpndnt_cplng}

\begin{figure}[h]
\centering
\includegraphics[scale=0.47]{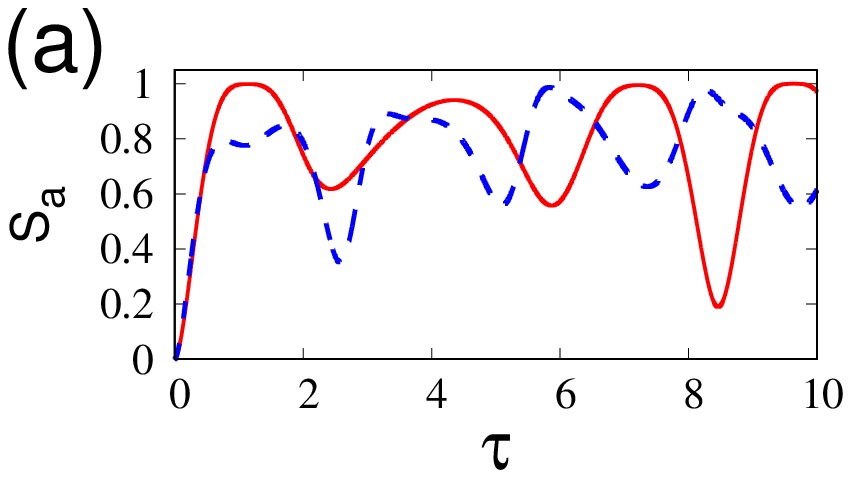}\hspace{0.5ex}
\includegraphics[scale=0.47]{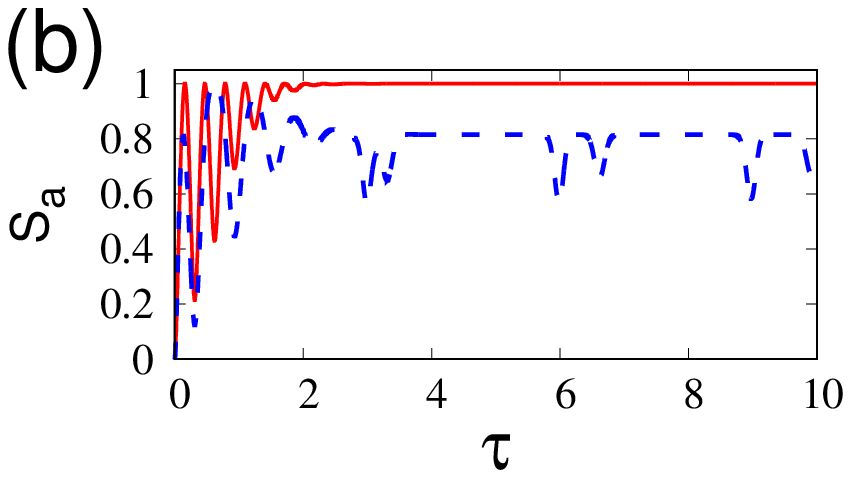}\\ \vspace{-7ex}
\includegraphics[scale=0.47]{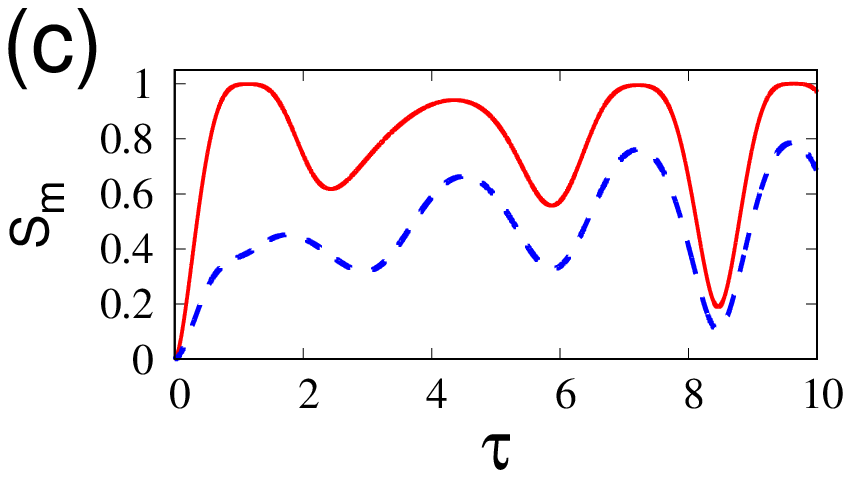}\hspace{0.5ex}
\includegraphics[scale=0.47]{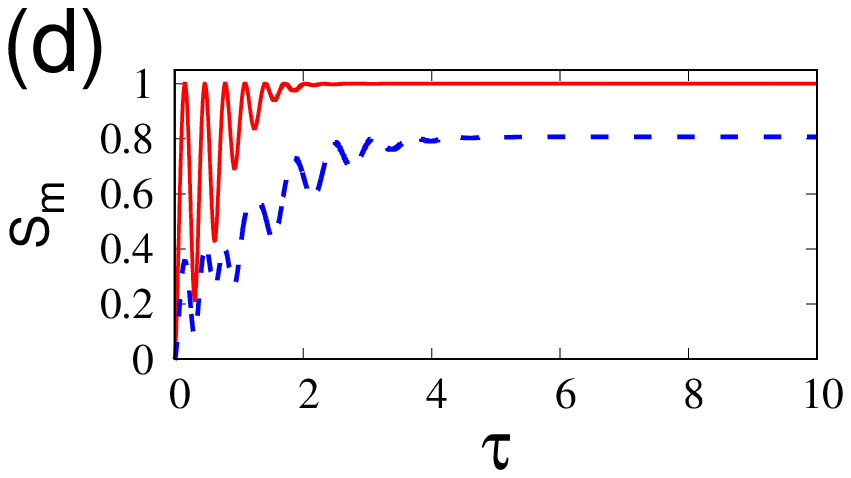}\vspace{-7ex}
\includegraphics[scale=0.47]{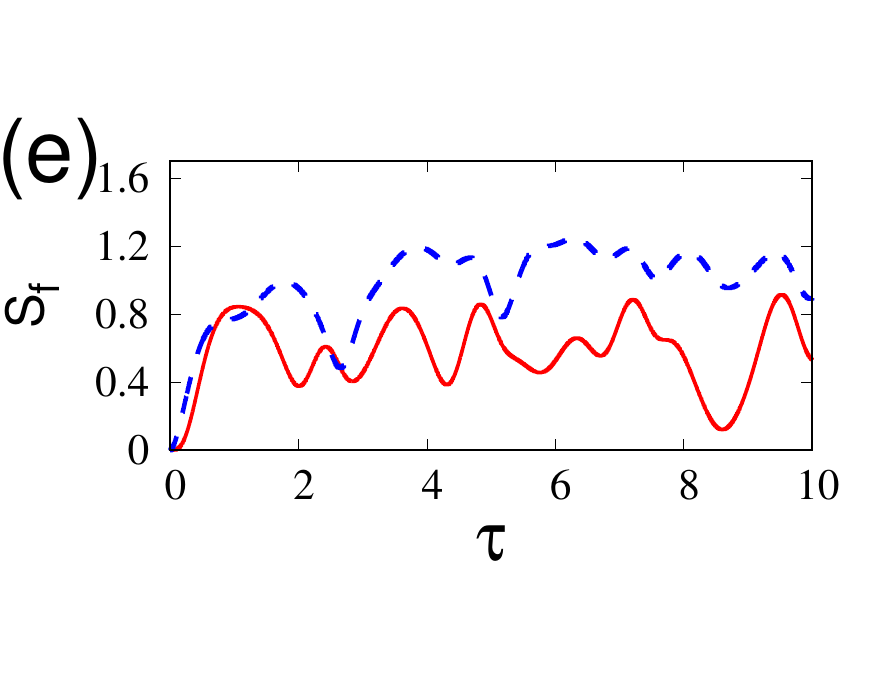}\hspace{0.5ex}
\includegraphics[scale=0.47]{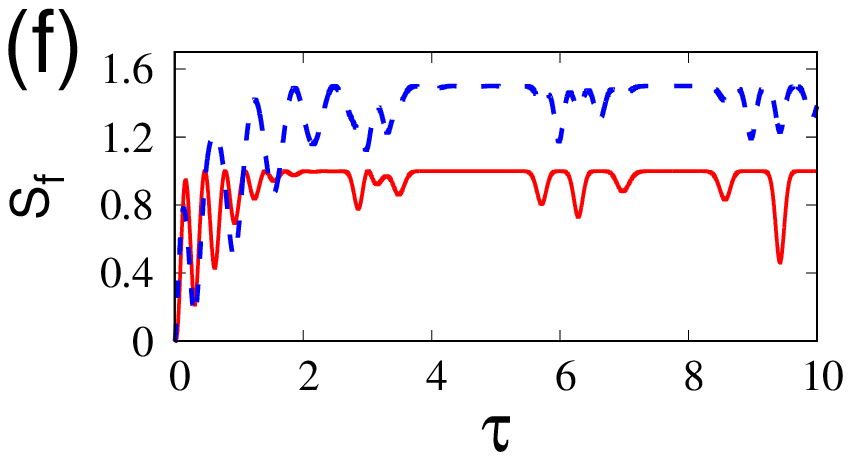}\vspace{-4ex}
\caption{S$_{\text{a}}$ (top panel), S$_{\text{m}}$ (centre panel) and  S$_{\text{f}}$ (bottom panel) vs $\tau$, for $\Omega = 10^{6}$ and $\phi =  \tfrac{1}{2} \pi$ (red), $\tfrac{1}{4} \pi$ (blue). $\alpha =1$ (first column) and $5$ (second column). }
\label{fig:param1_svne}
\end{figure}

\noindent 
We consider first an intensity-independent tripartite coupling, which  corresponds to  setting $f(N) = 1$. In figures \ref{fig:param1_svne}(a)-(f), the SVNE for the atom (S$_{\text{a}}$), mirror (S$_{\text{m}}$) and  field (S$_{\text{f}}$) are plotted as functions of $\tau$,  for $\Omega = 10^{6}$. Two points are noteworthy. First, for $\phi = \tfrac{1}{2} \pi$,  even for a small value of $\alpha$ (e.g., $\alpha =  1$),  S$_{\text{a}}$ equals unity at specific instants of time (figure \ref{fig:param1_svne}(a)). This is the maximum allowed value of the SVNE for a two-level atom. More importantly, with an increase in $\alpha$, this value remains constant over a long time interval (figure \ref{fig:param1_svne}(b)). It follows  from \eqref{eqn:soln_a}-\eqref{eqn:soln_c}, \eqref{eqn:red_den_mat_a} and \eqref{eqn:red_den_mat_m}  that for $\phi = \tfrac{1}{2} \pi$, the dynamics of S$_{\text{a}}$ and S$_{\text{m}}$ are identical. However, for other values of $\phi$ (e.g., $\tfrac{1}{4}\pi$), S$_{\text{a}}$ does not collapse to a constant value over a significant time interval for any value of $\alpha$ (figures \ref{fig:param1_svne}(a), (b)). In contrast,  the SVNE for the oscillator subsystem collapses to a non-zero value ($<1$) over a significant time interval for sufficiently large $\alpha$ and for $\phi=\tfrac{1}{4}\pi$ (figure \ref{fig:param1_svne}(d)). This allows for the  possibility of tuning the values of $\alpha$ and $\phi$ to retain entanglement collapse over long time intervals. This  could be potentially useful in quantum information transfer.

We have also verified that a further enhancement of the interval over which such collapses occur is possible if we consider an initial  field state $\ket{\alpha,m}$ ($m = 1,\, 2, \dotsc$), the  $m$-photon added coherent state instead of a standard CS.  $\ket{\alpha, \,m}$ is obtained~\cite{tara} by normalizing the state $(a^{\dagger})^{m}\ket{\alpha}$ to unity. The set $\{\ket{\alpha,\, m}\}$ provides a family  of states whose departure from coherence is precisely quantifiable. We have verified that increasing $m$ increases the collapse interval.

 No distinctive collapse in entanglement is seen in the field SVNE S$_{\text{f}}$ for any value of $\alpha$ and $\phi$. Unlike S$_{\text{a}}$ and S$_{\text{m}}$, the largest value of S$_{\text{f}}$ corresponds to  $\phi = \tfrac{1}{4} \pi$  (figures \ref{fig:param1_svne}(e)-(f)). 

\begin{figure}
\centering
\includegraphics[scale=0.47]{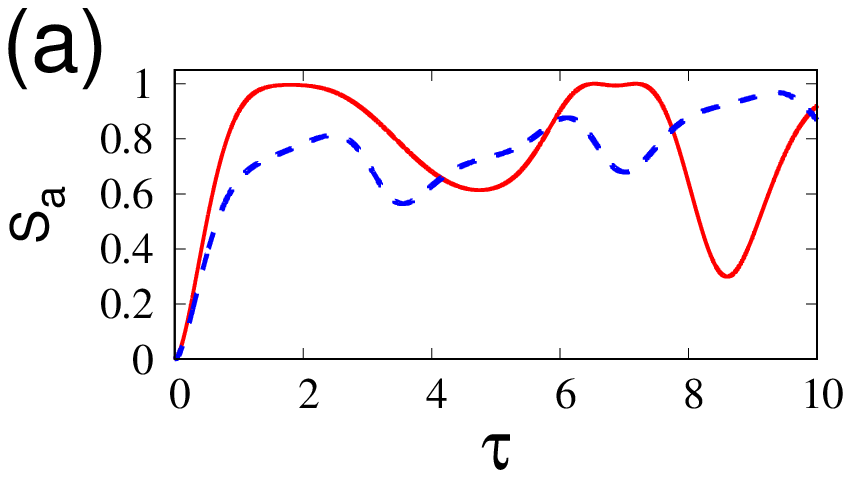}\hspace{0.5ex}
\includegraphics[scale=0.47]{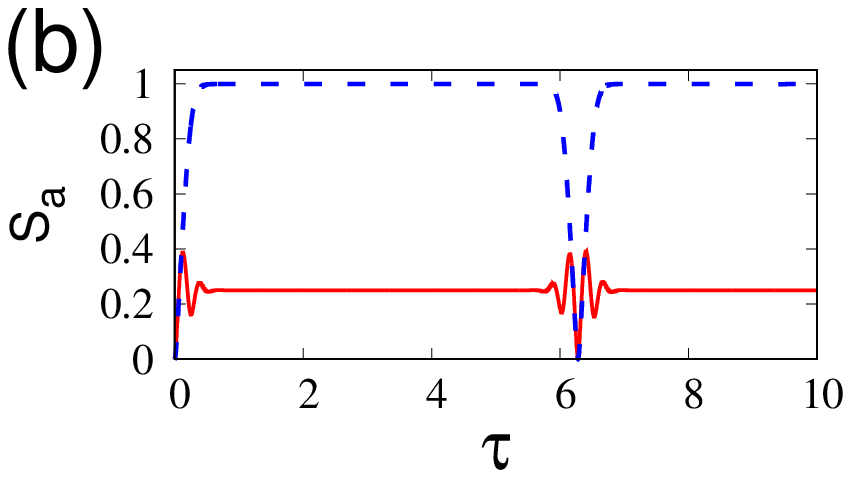}\\ \vspace{-7ex}
\includegraphics[scale=0.47]{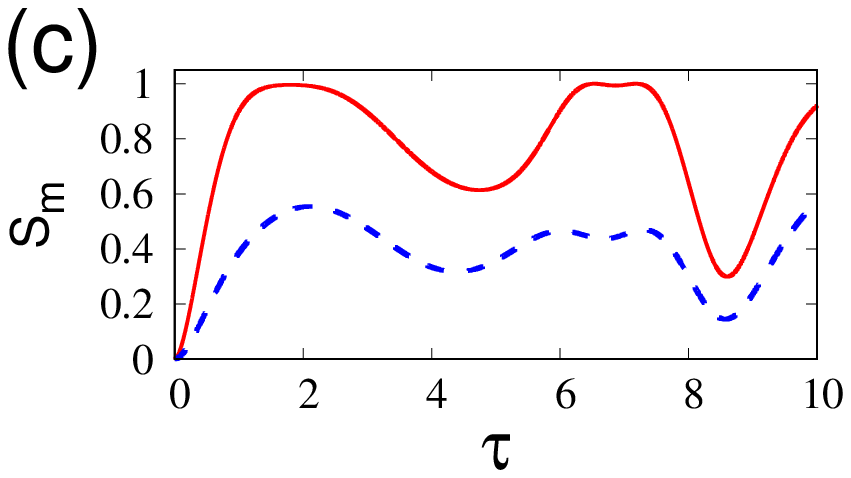}\hspace{0.5ex}
\includegraphics[scale=0.47]{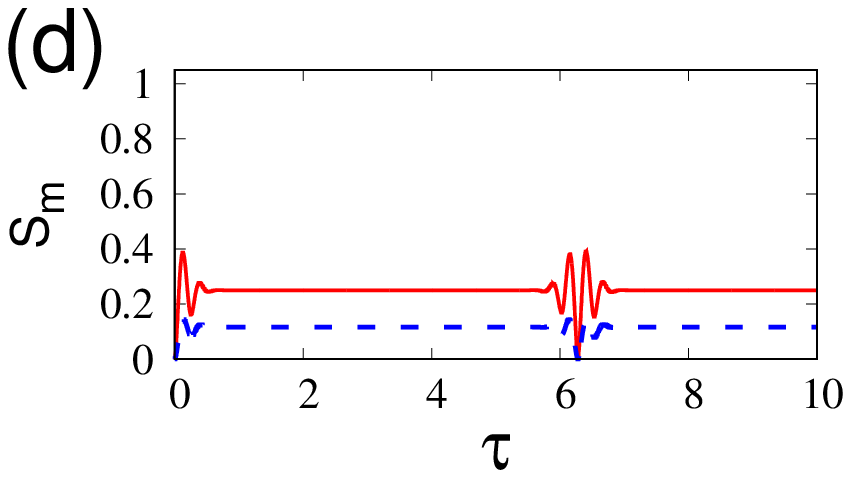}\\ \vspace{-7ex}
\includegraphics[scale=0.47]{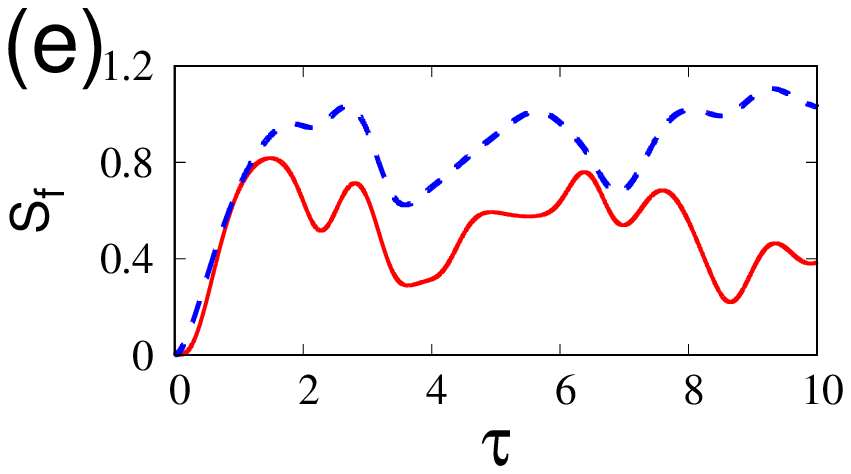}\hspace{0.5ex}
\includegraphics[scale=0.47]{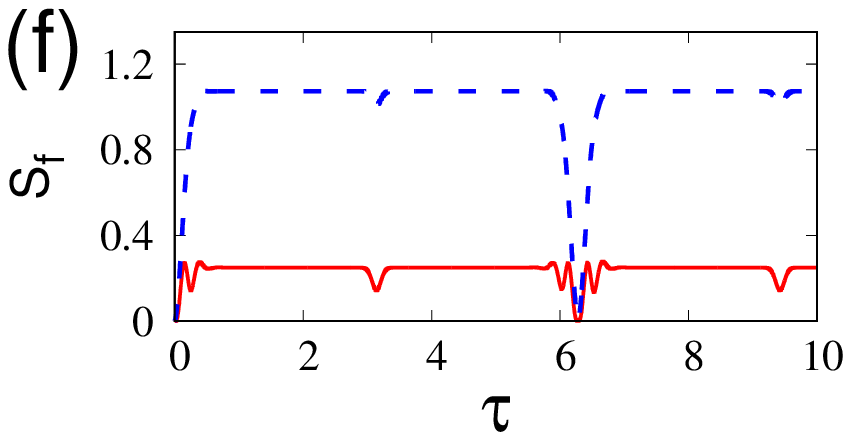}\vspace{-4ex}
\caption{S$_{\text{a}}$ (top panel), S$_{\text{m}}$ (centre panel) and  S$_{\text{f}}$ (bottom panel) vs $\tau$, for $\Omega = 10^{6}/\sqrt{2}$ and $\phi =  \tfrac{1}{2} \pi$ (red), $\tfrac{1}{4} \pi$ (blue). $\alpha =1$ (first column) and $5$ (second column). }
\label{fig:param2_svne}
\end{figure}

In contrast to the foregoing, when $\Omega$ is reduced to the value $10^{6}/\sqrt{2}$, relatively long-time-interval collapses of the SVNE to a non-zero value are observed  for sufficiently large values of $\alpha$ for both $\phi = \tfrac{1}{2} \pi$ and $\tfrac{1}{4} \pi$ (figure \ref{fig:param2_svne}(b),(d),(f)). For low values of $\alpha$ this feature is absent although relatively high values of SVNE are reached at specific instants of time.

\begin{figure}[h]
\centering
\includegraphics[scale=0.47]{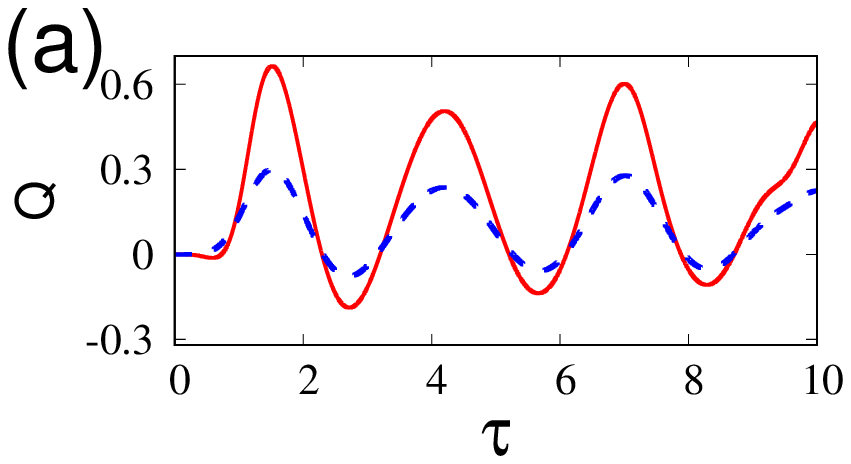}\hspace{1ex}
\includegraphics[scale=0.47]{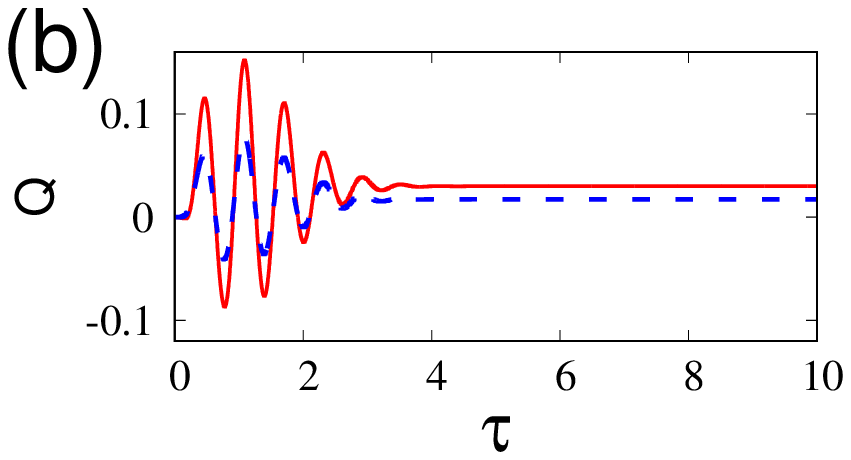}\\ \vspace{-7ex}
\includegraphics[scale=0.47]{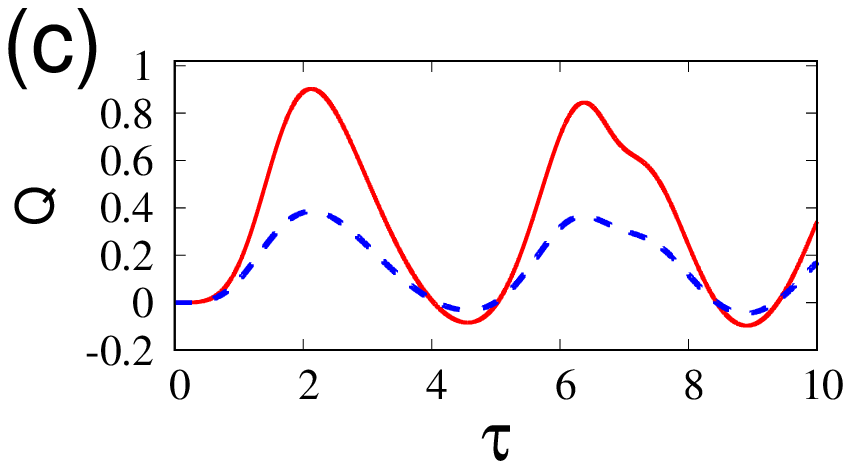}\hspace{1ex}
\includegraphics[scale=0.47]{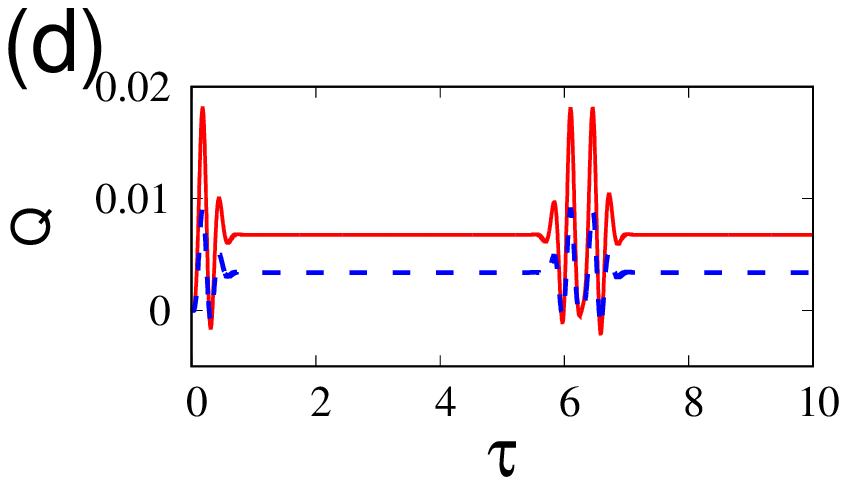}\\ \vspace{-4ex}
\vspace{-0ex}
\caption{Mandel Q parameter vs  $\tau$ for $\phi =  \tfrac{1}{2} \pi$ (red), $\tfrac{1}{4} \pi$ (blue).   $\alpha = 1$ (first column) and $5$ (second column).  $\Omega = 10^{6}$ (top panel) and $10^{6}/\sqrt{2}$ (bottom panel).}
\label{fig:manq_k_0}
\end{figure}

 The Mandel Q parameter for the field subsystem displays collapse over a long time interval for both $\Omega = 10^{6}$ and $\Omega = 10^{6}/\sqrt{2}$ for sufficiently high $\alpha$ (figures \ref{fig:manq_k_0}(b), (d)). Such collapses are not present for small $\alpha$ (figures \ref{fig:manq_k_0}(a), (c)). We  see sub-Poissonian signatures (negative values) of Q in all cases. However, for smaller $\alpha$, these are  more prominent. We have verified that these features are also reflected in the dynamics of  the mean photon number, its variance and the atomic inversion parameter.

%%%%%%%%%%%%%%%%%%%%%%%%%%%%%%%%%%%%%%%%%%%%%%%%%%%%%%%%%%%%%%%%%%%%%%%%%%
\subsection{Intensity-dependent tripartite coupling}
\label{intn_dpndnt_cplng}
The results presented in the foregoing enable us to infer  that entanglement collapses   of SVNE to non-zero values over a long time interval  are present  only for sufficiently large values of $\alpha$. We now examine how various forms of IDC  affect the collapse intervals of S$_{\text{a}}$ and S$_{\text{m}}$.

\begin{figure*}%[h]
 \centering
 \includegraphics[height=3.8cm, width=5.9cm]{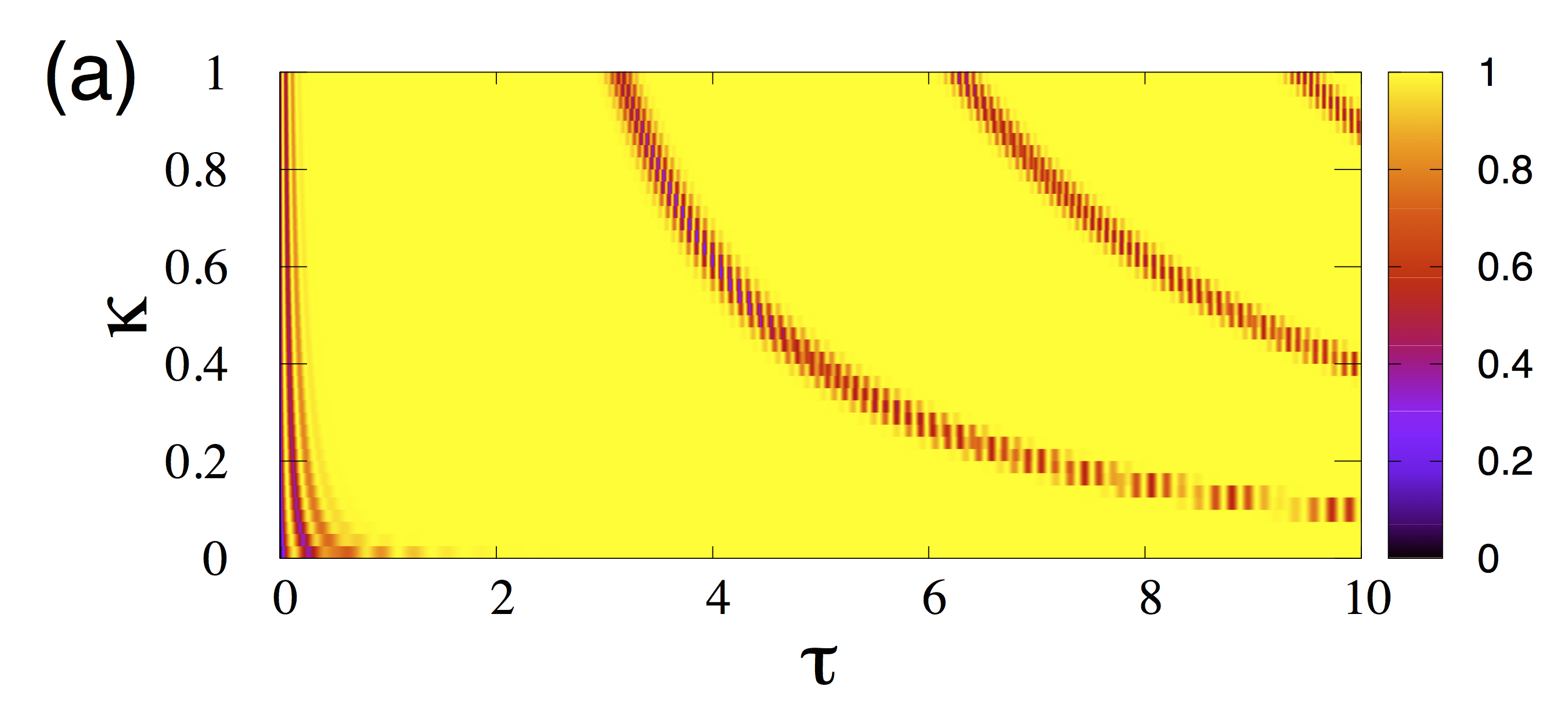}\vspace{-1ex}
 \includegraphics[height=3.8cm, width=5.9cm]{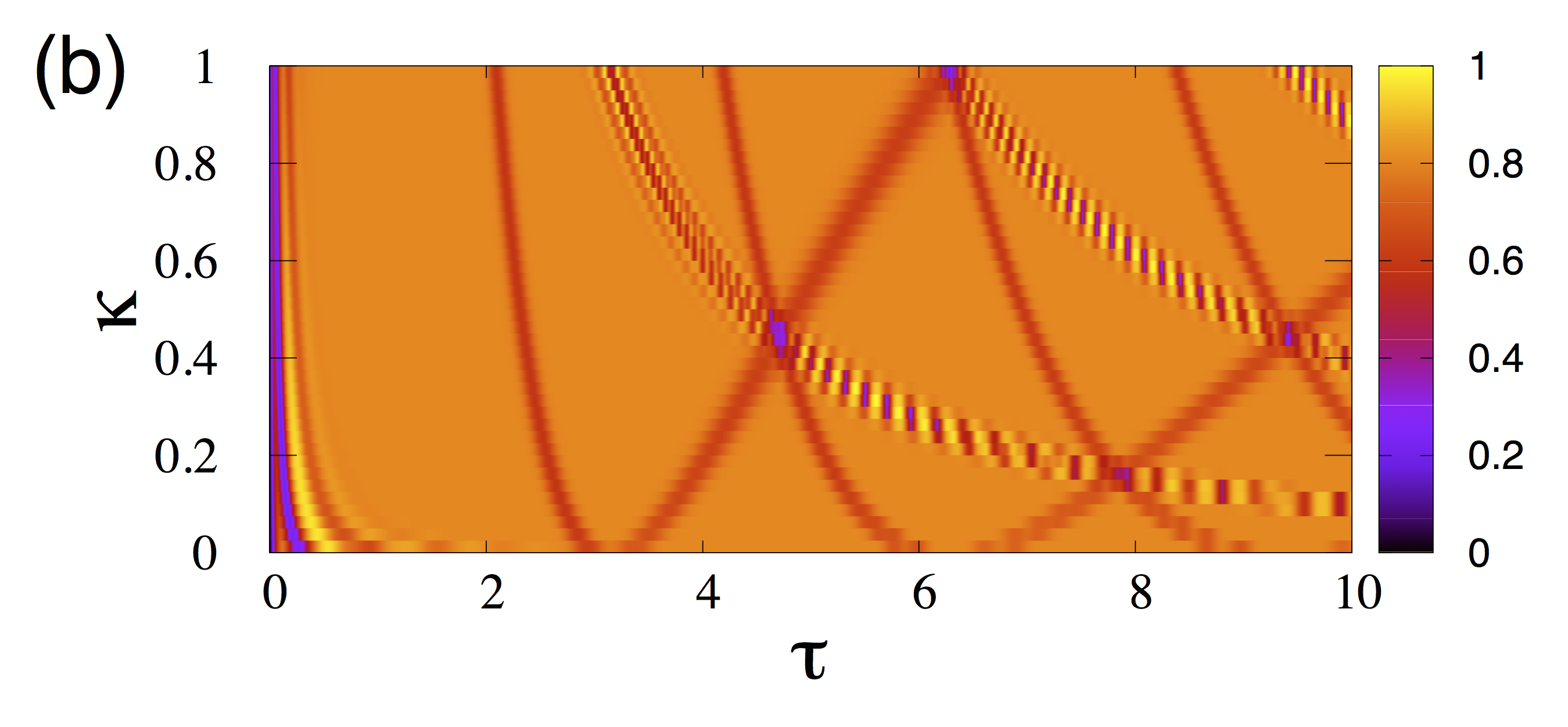}\vspace{-1ex}
 \includegraphics[height=3.8cm, width=5.9cm]{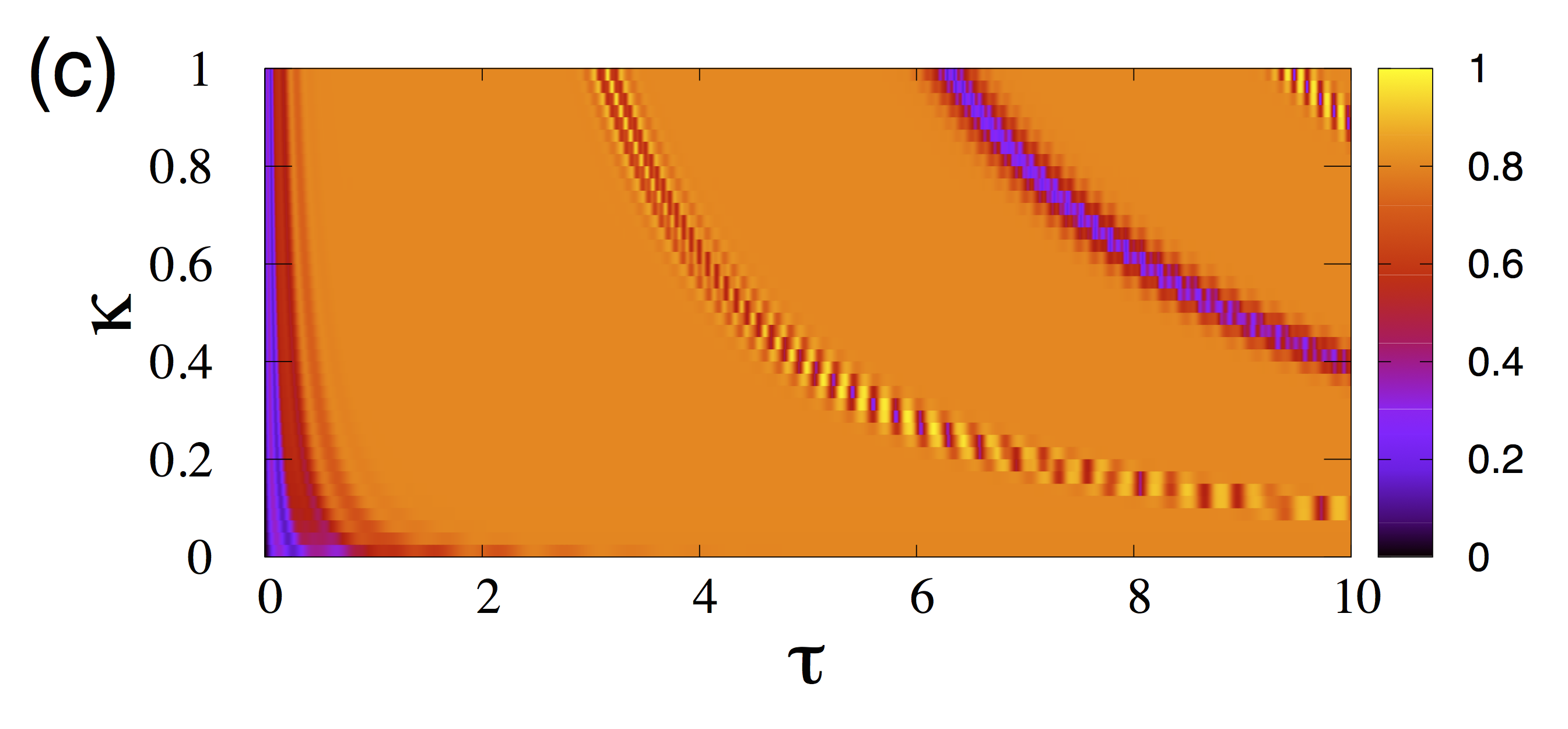}
 \vspace{-0.5ex}
 \caption{Contour plots of SVNE vs $\tau$ and $\kappa$ with  $\alpha = 8$ and  $\Omega = 10^{6}$,  for (a) the atom and the oscillator for $\phi = \tfrac{1}{2} \pi$;  (b) the atom for $\phi = \tfrac{1}{4}\pi$; (c)   the  oscillator for $\phi = \tfrac{1}{4} \pi$.}
 \label{fig:param1_svne_3d_kappa_loop}
\end{figure*}

We find that, when $f(N) = (a^{\dagger}a)^{-1/2}$ collapses  are destroyed, while for $f(N) = (a^{\dagger}a)^{1/2}$ the collapse intervals are reduced, compared to the intensity-independent situation. For $f(N) = (1 + \kappa\, a^{\dagger}a)^{1/2}$  ($0 < \kappa \le 1)$,  an increase in $\kappa$ decreases the interval of  collapse,  but increases the number of collapses within a given time interval.  Figures \ref{fig:param1_svne_3d_kappa_loop}(a)-(c) show contour plots of the SVNE as a function of  $\tau$  and the IDC parameter $\kappa$.

We now compare these results with those obtained for a $\Lambda$ atom interacting with a coupling field and a probe field \cite{laha1}. In both systems, the SVNE  collapses  to  a constant non-zero value over a significant interval of time in the absence of intensity-dependent field-atom coupling, and  the dynamics is very sensitive to the value of the parameter $\kappa$. In the latter  system,  an increase in $\kappa$ produced a spectacular bifurcation cascade in the qualitative behaviour of the SVNE.  While such remarkable changes do not appear in the present model,  there are still distinctive features in the SVNE that are controlled by the  precise value of $\kappa$ (figures \ref{fig:param1_svne_3d_kappa_loop}(a)-(c)). Additionally, in the case at hand,  the SVNE collapses to its maximum possible value,  as explained earlier. 

Finally, we observe in passing that for $f(N) = (a^{\dagger} a)^{-1/2}$ (and  $\Omega = 10^{6}$), the Q parameter for the field does not become negative over the time interval considered  for any value of $\alpha$ and $\phi$, in contrast to its behaviour for $f(N)= (a^{\dagger} a)^{1/2}$ or $(1+ \kappa\, a^{\dagger} a)^{1/2}$.
\begin{figure*}
\centering
\includegraphics[scale=0.42, angle = -0]{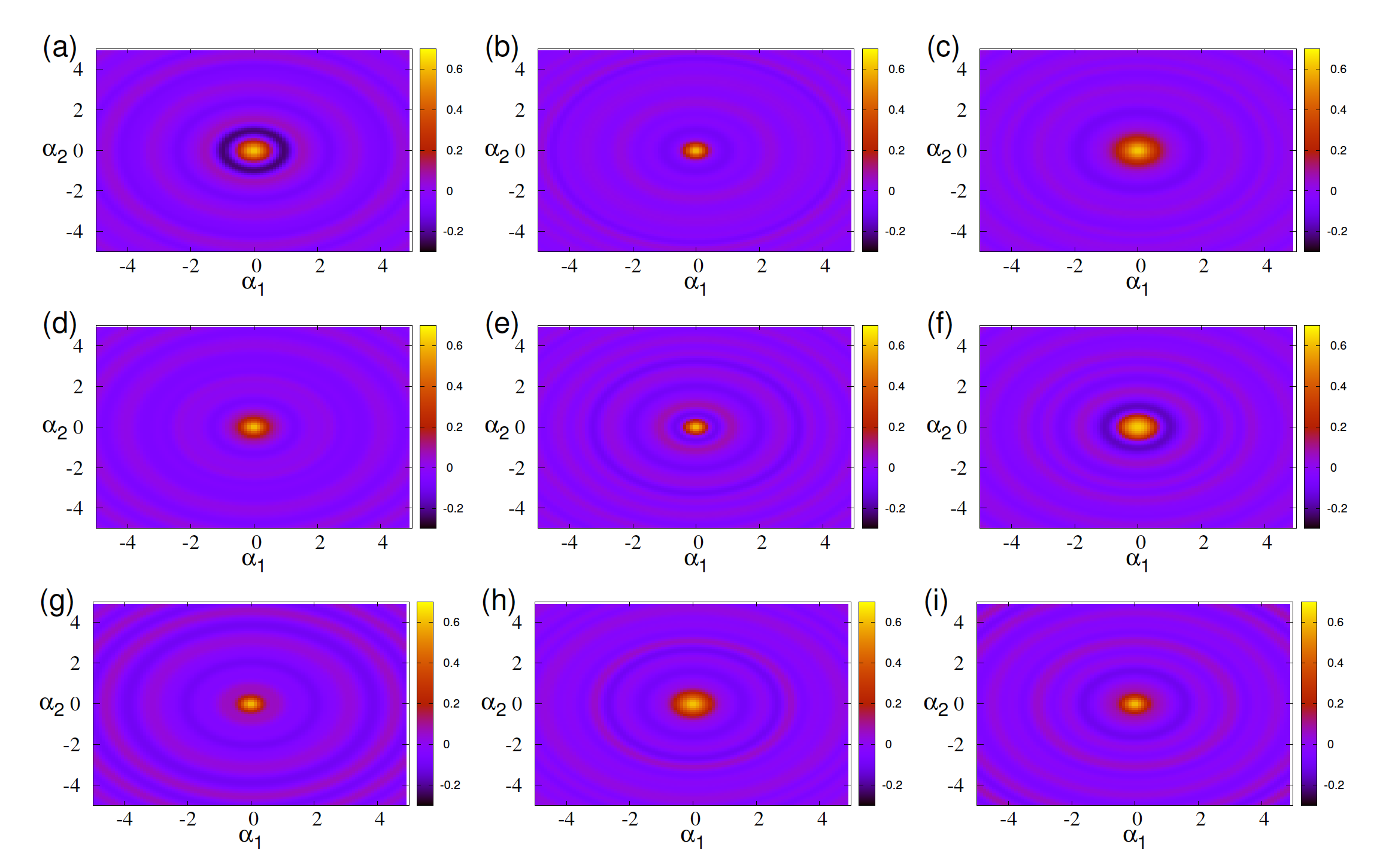}\hspace{1ex}
\vspace{-1ex}
\caption{$W_{f}(\alpha)$ vs $\alpha_{1}$ and $\alpha_{2}$ for $\phi = \tfrac{1}{2}\pi$, $\Omega = 10^{6}$ and $f(N) = (1 + \kappa\, a^{\dagger} a)^{1/2}$ for $\kappa = 0$ (top panel), $0.5$ (middle panel) and $1$ (bottom panel)  at instants $\tau= 2$ (first column), $5$ (second column) and $8$ (third column) respectively.}
\label{cavity_wigner2}
\end{figure*}

%%%%%%%%%%%%%%%%%%%%%%%%
\subsection{Wigner functions of the field}
The  nonclassical nature of the field at various instants of time is reflected in the negativity of the Wigner functions at those instants. We have derived  an expression for the Wigner distribution $W_{f}(\alpha)$  ($\alpha = \alpha_{1} + i \,\alpha_{2}$) in Appendix \ref{appendix_wigner_function}. For the generic state given in \eqref{eqn:final_state}, we have
 \begin{align}
  W_{f}(\alpha) &= \frac{2}{\pi} \sum_{n,k,l=0}^{\infty} (-1)^{n}\, q_{k}\, q_{l}^{*} \big[ A_{k}\, A_{l}^{*}  + B_{k}\, B_{l}^{*} \big]  \times \nonumber \\
                        &\hspace{4cm} \expect{n}{D^{\dagger}}{k} \expect{l}{D^\dagger}{n} \nonumber \\
                        &+ \frac{2}{\pi} \sum_{n,k,l=0}^{\infty}(-1)^{n}\, q_{k}\, q_{l}^{*} C_{k}\, C_{l}^{*} \times \nonumber \\
                        &\hspace{2cm} \expect{n}{D^{\dagger}}{k-1} \expect{l-1}{D^{\dagger}}{n},
\label{eqn:wigner_cav}
\end{align}
with the identification 
\begin{equation}
 \expect{n}{D}{m} = 
  \left\{
    \begin{array}{l}
      e^{-\frac{1}{2}|\alpha|^{2}}\, \sqrt{\frac{m!}{n!}}\,\, \alpha^{n-m} \, L_{m}^{n-m}(|\alpha|^{2}) \;\; (n\ge m)\\[6pt]
      e^{-\frac{1}{2}|\alpha|^{2}}\, \sqrt{\frac{n!}{m!}}\, \,(-\alpha^{*})^{m-n} \, L_{n}^{m-n}(|\alpha|^{2})\;\; (n<m) 
    \end{array}
  \right.
\end{equation}
 where $L_{n}^{m}(x)$ is the associated Laguerre polynomial. Here $A_{i}, \,B_{i}$ and $C_{i}$ are functions of both time and $f(N)$.

%For $f(N) = 1$, the negative regions in the Wigner distribution become more pronounced with elapse of time (figures \ref{cavity_wigner2}(a)-(c)). 
For completeness, we plot the Wigner functions for different values of $\tau$ and $\kappa$ in figures  \ref{cavity_wigner2}(a)-(i).
%In the case  $f(N) =  (1 + \kappa\, a^{\dagger} a)^{1/2}$, the  Wigner functions for different values of $\kappa$ and $\tau$  are shown  in figures \ref{cavity_wigner2}(d)-(i). 
 It is evident that  for a fixed value of $\kappa$ , negativity of the Wigner function does not necessarily increase with time (see for instance, figures \ref{cavity_wigner2}(d)-(f)).
 
%%%%%%%%%%%%%%%%%%%%%%%%%%%%%%%%%%%%%%%%%%%%%%%%%%%%%%%%%%%%%%%%%%%%%%%%%%

\section{The optical tomogram and squeezing properties of the field}  
\label{tomo_section}

\subsection{The optical tomogram} 

We now examine the  squeezing of the state of the field subsystem as the full system evolves in time. It is useful to carry out this analysis in terms of the optical tomogram $\omega(X_{\theta}, \theta)$ of the field, because the nonclassical properties of the field are conveniently  reflected in this quantity: $\omega(X_{\theta}, \theta)$ is just the Radon transform of the Wigner function $W(p, q)$ 
derived  in Appendix \ref{appendix_wigner_function}. It is defined as  
 \begin{equation}
  \omega(X_{\theta}, \theta) = \int \delta(\mathbb{X}_{\theta} - X_{\theta} \mathbb{I}) \, W(p, q) \, \frac{dq\, dp}{2 \pi},
  \label{eqn:rad_wig}
 \end{equation}
where $\mathbb{I}$ is the identity operator, and 
\begin{equation}
 \mathbb{X}_{\theta} = \hat{q}\cos\theta + \hat{p} \sin\theta = (a e^{-i \theta} + a^\dagger e^{i \theta})/\sqrt{2}
\label{eqn:xtheta_op}
\end{equation}
is the homodyne quadrature operator expressed in terms of the photon destruction and creation operators. We have $ \mathbb{X}_{\theta} \ket{X_{\theta}, \theta} = X_{\theta} \ket{X_{\theta}, \theta}$,  where 
\begin{equation}
  \ket{X_{\theta}, \theta} = \frac{1}{\sqrt{\pi}} \exp\,\big(\!-\tfrac{1}{2}X_{\theta}^{2}  
  - \tfrac{1}{2} e^{i 2 \theta}a^{\dagger 2}+ \sqrt{2} e^{i \theta} X_{\theta}a^{\dagger}\big) \ket{0}.
  \label{eqn:ket_x_theta1}
 \end{equation} 
 We make use of the fact that $\omega(X_{\theta}, \theta) =  \expect{X_{\theta}, \theta}{\rho_{f}}{X_{\theta}, \theta}$. It follows that, for a state $\ket{\psi} = \sum_{n=0}^{\infty} c_{n} \ket{n}$, 
\begin{equation}
 \omega(X_{\theta}, \theta) = \frac{e^{-X_{\theta}^2}}{\sqrt{\pi}} \Bigg| \sum_{n=0}^{\infty} c_{n} \frac{ e^{-i n \theta}}{\sqrt{2^{n}\: n!}} H_{n}(X_{\theta})\Bigg|^2.
 \label{eqn:w_cnm}
\end{equation}
($H_{n}$ is the Hermite polynomial of order $n$.)  In what follows, \eqref{eqn:w_cnm} will be used extensively for numerically estimating the squeezing properties of the field from the tomogram.
 
\begin{figure}[h]
 \centering
 \includegraphics[scale=0.1, angle = -0]{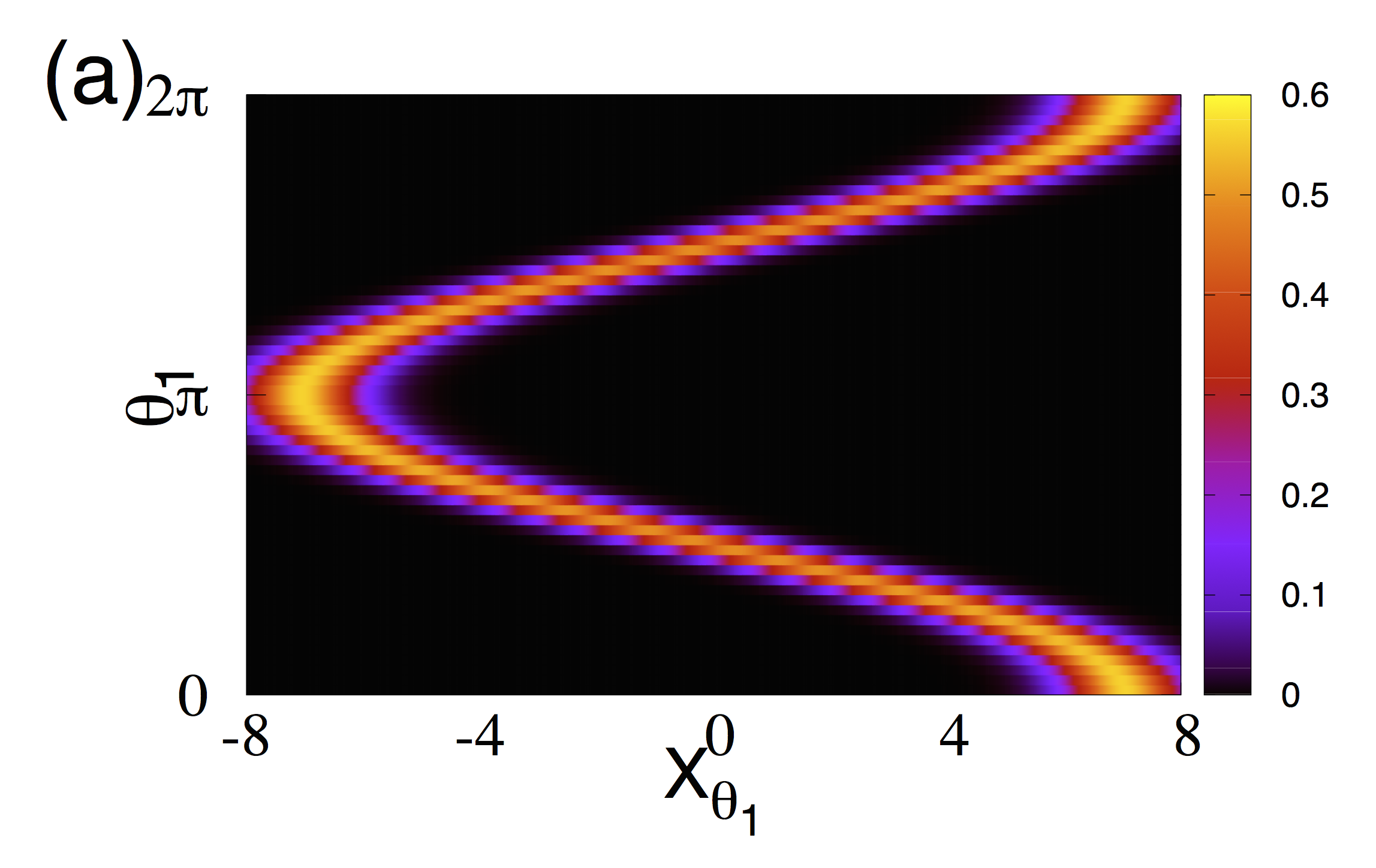}
 \includegraphics[scale=0.1, angle = -0]{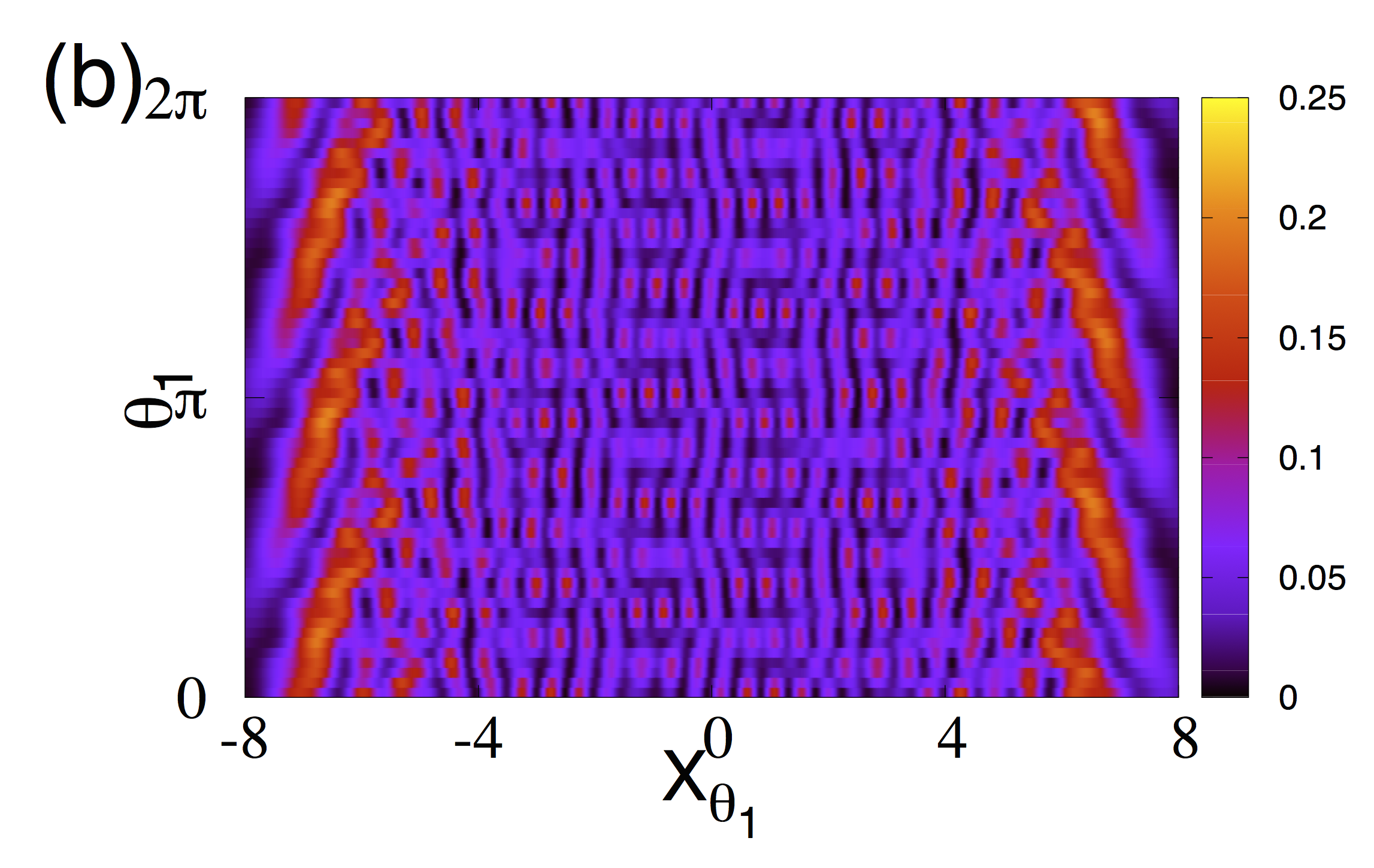}
 \includegraphics[scale=0.1, angle = -0]{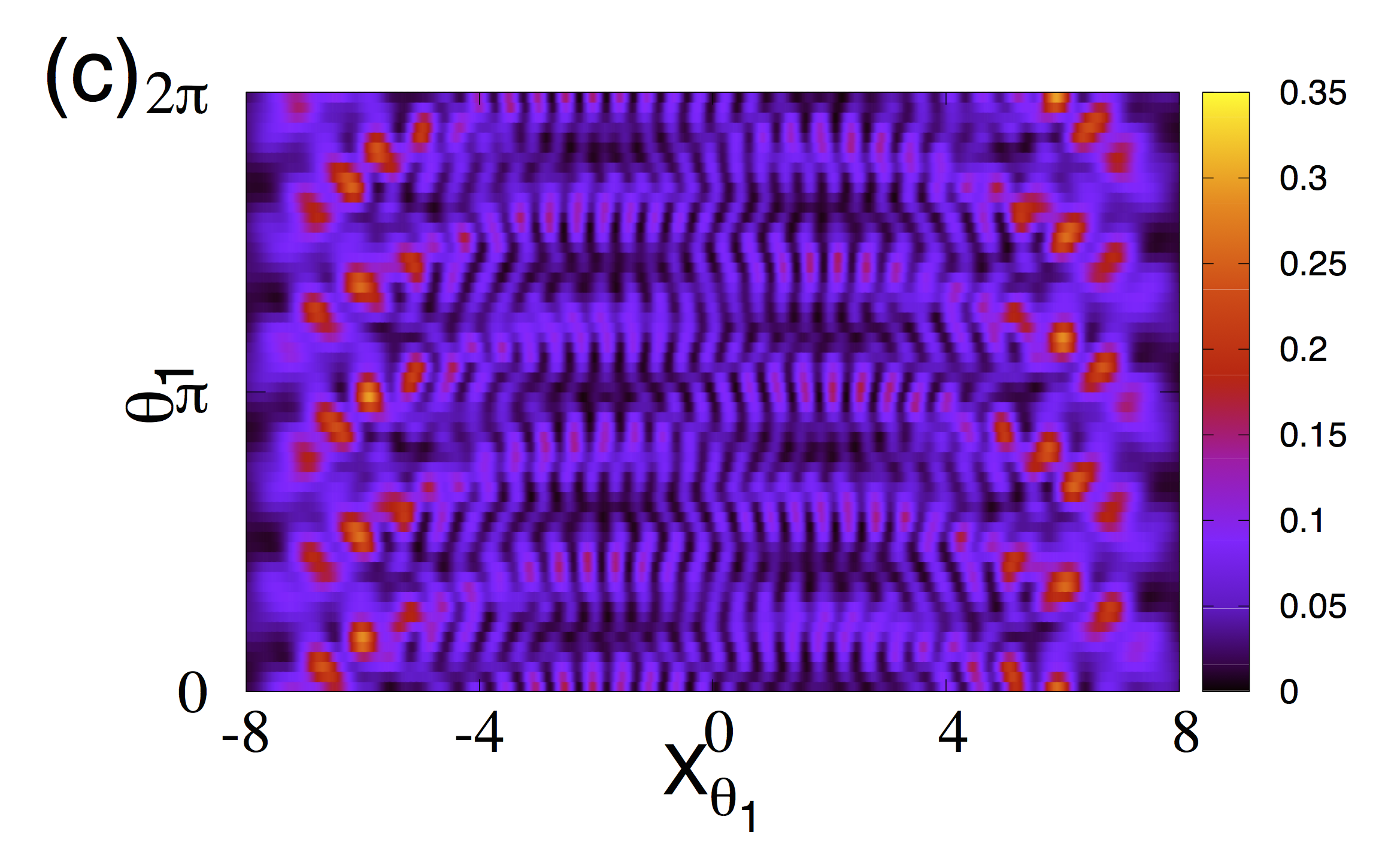}
 \includegraphics[scale=0.1, angle = -0]{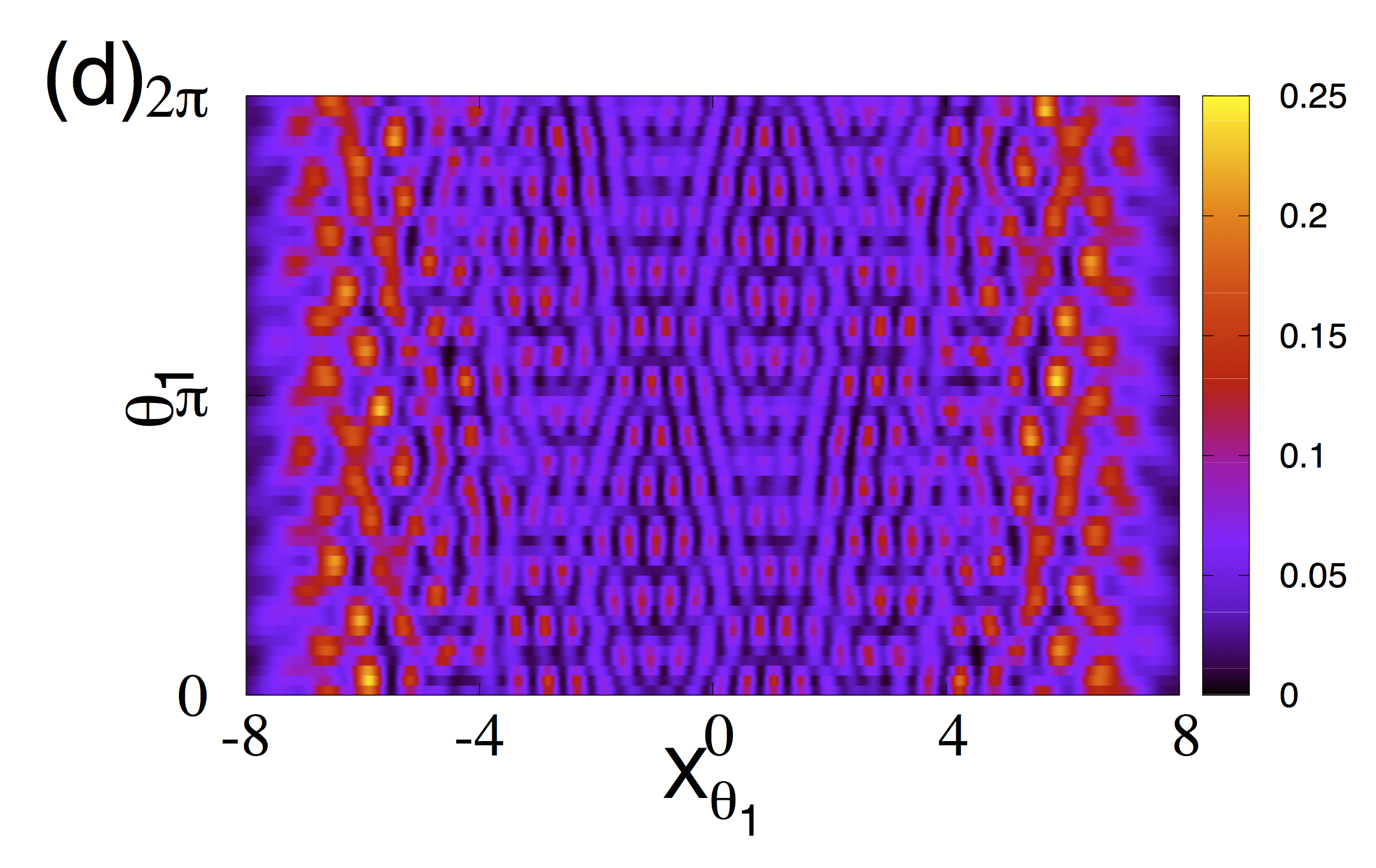}
\vspace{-1ex}
\caption{Field tomograms  at  (a) $\tau=  0$, (b) $\tau =  2$,  (c) $\tau = 5$ and (d) $\tau = 8$, for $\Omega = 10^{6}, \,\phi = \tfrac{1}{2}\pi, \,\alpha=5, \,\kappa = 0$ for an initial CS.}
\label{fig:cavity_tomo}
\end{figure}

The generic tomogram is pictorially represented as an intensity  plot of $\omega(X_{\theta}, \theta)$ versus $X_{\theta}$ and $\theta$. In figure \ref{fig:cavity_tomo}(a) we present the tomogram of an initial CS with $\alpha = 5$. In contrast to this, the tomograms at later times are significantly more complex in appearance: this is  illustrated, for the  sake of completeness, in figures \ref{fig:cavity_tomo}(b-d). These tomograms correspond to the Wigner functions at the instants specified in figures \ref{cavity_wigner2}(a-c).

%%%%%%%%%%%%%%%%%%%%%%%%%%%%%%%%%%%%%%%%%%%%%%%%%%%%%%%%%%%%%%%%%%%%%%%%%%%
\subsection{Entropic and quadrature squeezing}

The tomographic entropy $S(\theta)$ for a subsystem, defined as
 \begin{equation}
  S(\theta) =  - \int \omega(X_{\theta}, \theta)\,\ln\,[\omega(X_{\theta}, \theta)]\, dX_{\theta},
  \label{eqn:single_mode_tomo_ent}
 \end{equation}
 satisfies the entropic uncertainty relation (EUR) \cite{orlowski}
 \begin{equation}
   S(\theta) + S(\theta+\tfrac{1}{2}\pi) \geq \ln\, (\pi e)
  \label{eqn:eur_single_mode}
 \end{equation}
 at every instant of time. A state with entropy in either quadrature ($\theta$ or $\theta + \tfrac{1}{2} \pi$) less than $\tfrac{1}{2} \ln\,(\pi e)$  displays entropic squeezing in that quadrature. The optical tomogram $\omega(X_{\theta}, \theta)$ is non-negative, and satisfies $  \int \omega(X_{\theta}, \theta) \,dX_{\theta}= 1$.   Hence, we can calculate the moments of the quadrature operators from $\omega(X_{\theta}, \theta)$ in a straightforward manner. For any specific value of $\theta$, we have 
  \begin{equation}
   \aver{X^{n}} = \int  X^{n}\, \omega(X_{\theta}, \theta)\, dX_{\theta}.
   \label{eqn:tomo_moments}
 \end{equation}
 
\begin{figure}[h]
\centering
\includegraphics[scale=0.48]{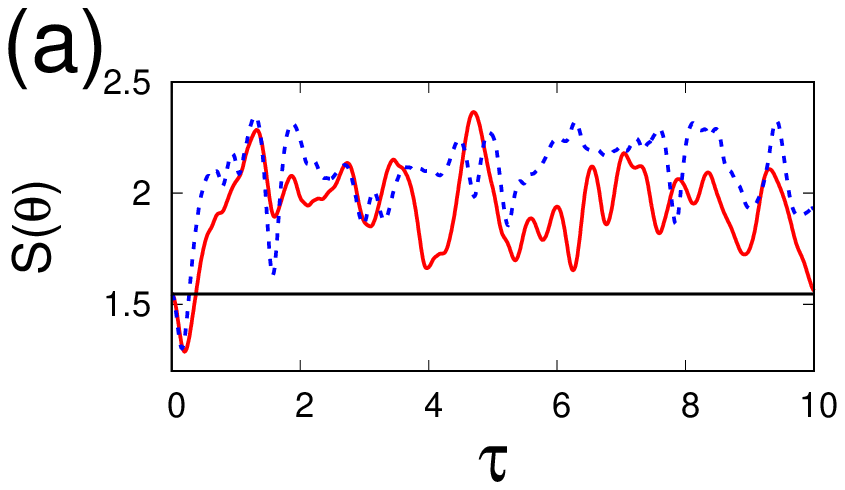}
\includegraphics[scale=0.48]{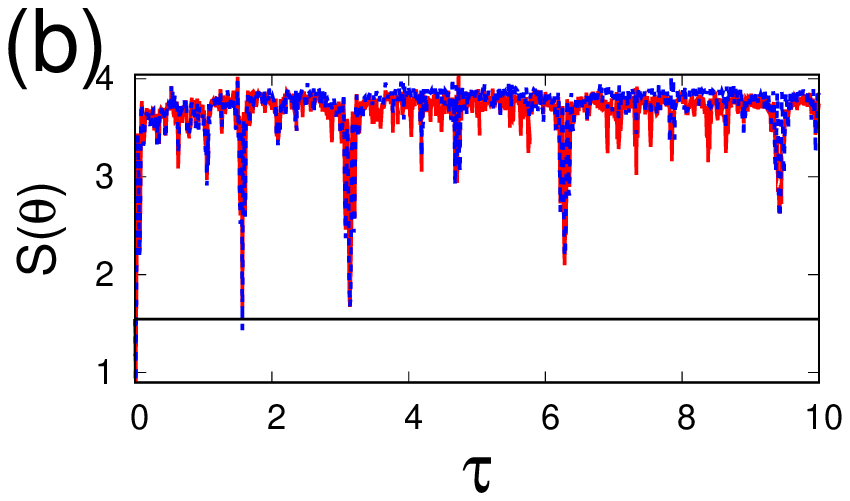}\\ \vspace{-6ex}
\includegraphics[scale=0.48]{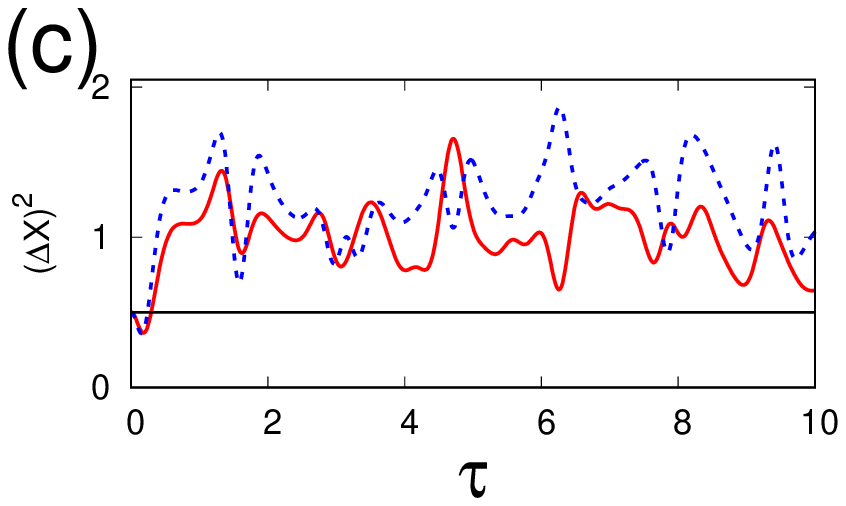}
\includegraphics[scale=0.48]{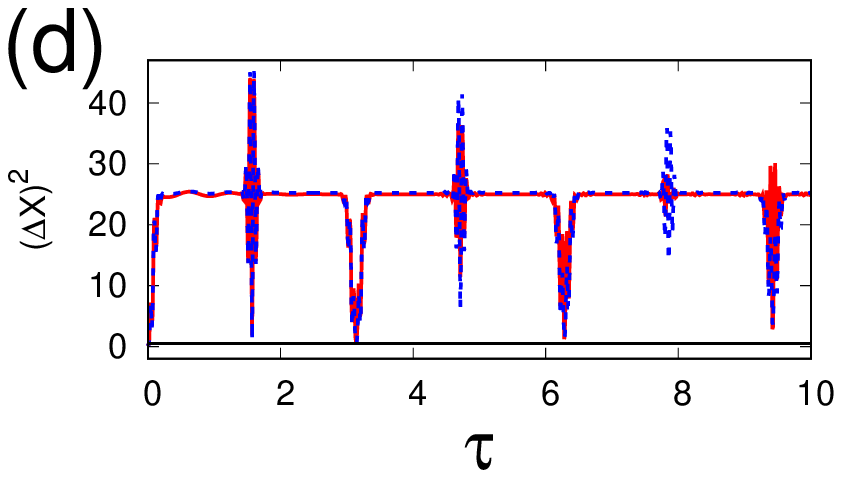}
\vspace{-5ex}
\caption{$S(\theta)$ (top panel) and $(\Delta X)^{2}$ (bottom panel) vs $\tau$ for $\Omega = 10^{6}$, $\theta = 0$, $\phi =  \tfrac{1}{2} \pi$ (red) and $\phi =  \tfrac{1}{4} \pi$ (blue).  $\alpha = 1$ (first column) and $5$ (second column). The horizontal black line is the reference value below which squeezing occurs.}
\label{fig:ent_k_0_param1}
\end{figure}

\begin{figure}[h]
\centering
\includegraphics[scale=0.48]{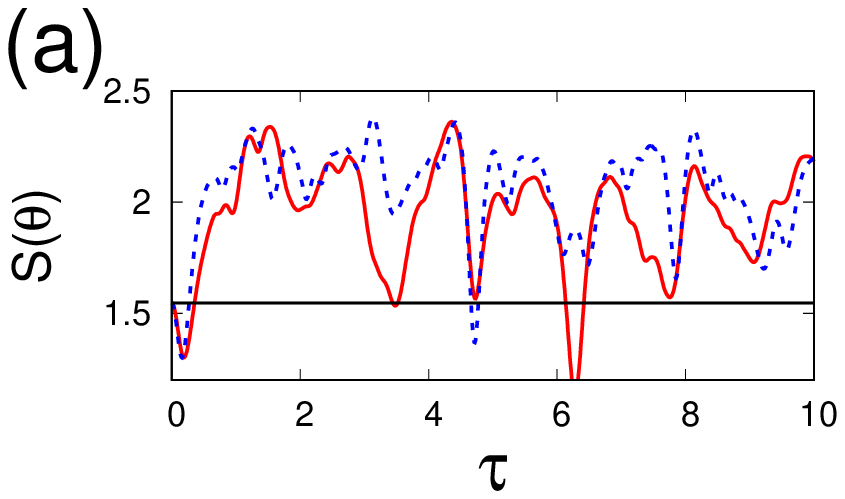}
\includegraphics[scale=0.48]{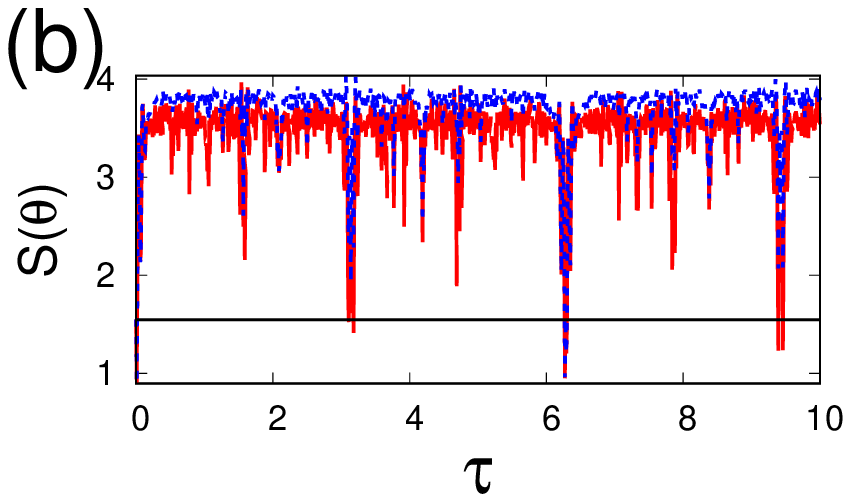}
\vspace{-5ex}
\caption{$S(\theta)$ vs $\tau$ for $\Omega = 10^{6}/\sqrt{2}$, $\theta = 0$, $\phi = \tfrac{1}{2} \pi$ (red) and $\phi =  \tfrac{1}{4} \pi$ (blue). (a) $\alpha =1$,  (b) $\alpha =  5$.  The  horizontal black line is the reference value below which squeezing occurs.}
\label{fig:ent_k_0_param2}
\end{figure}

 For $\Omega = 10^{6}$ and  $\theta = 0$, for instance, the tomographic entropy is squeezed  for $\alpha = 1$ although only at a few instants (figure \ref{fig:ent_k_0_param1}(a)). With an increase in the value of $\alpha$,  the extent of squeezing does not significantly change (compare figures \ref{fig:ent_k_0_param1}(a), (b)). Variances in quadrature operators are also not significantly squeezed for these values of the parameters, except at a few instants (figures \ref{fig:ent_k_0_param1}(c), (d)). In contrast, for $\Omega = 10^{6}/\sqrt{2}$,  the tomographic entropy exhibits more squeezing during the time interval considered (figures \ref{fig:ent_k_0_param2}(a)-(b)).

For completeness,  we report that over the same interval of time ($0\le \tau \le 10$) and  $f(N) = (a^{\dagger} a)^{1/2}$, the state displays entropic squeezing more frequently for $\Omega = 10^{6}$ and less frequently for $\Omega = 10^{6}/\sqrt{2}$ as $\alpha$ is increased.  For $f(N) = (a^{\dagger} a)^{-1/2}$, the frequency with which squeezing occurs 
increases with increasing $\alpha$ for both values of $\Omega$.

\begin{figure}[h]%\vspace{-2ex}
\centering
\includegraphics[scale=0.48]{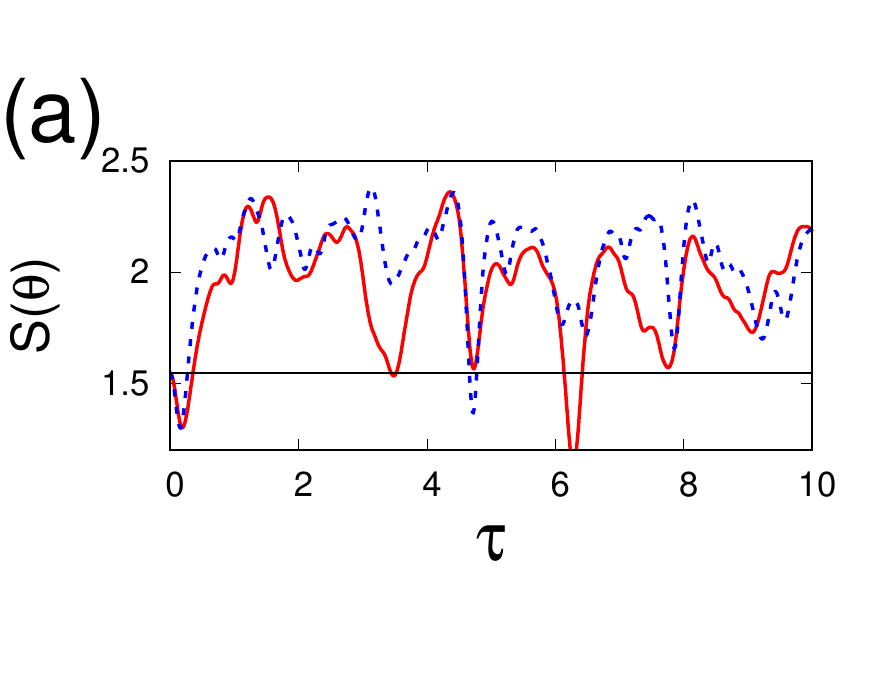}
\includegraphics[scale=0.48]{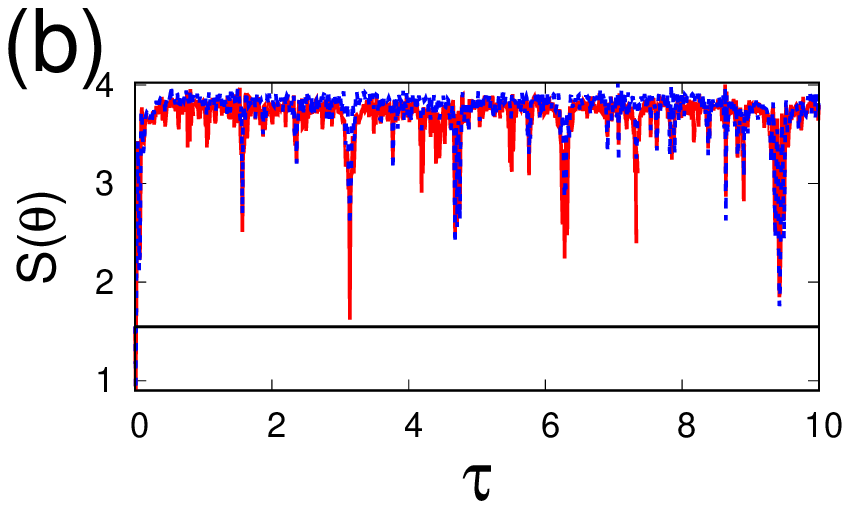}\\ \vspace{-6ex}
\includegraphics[scale=0.48]{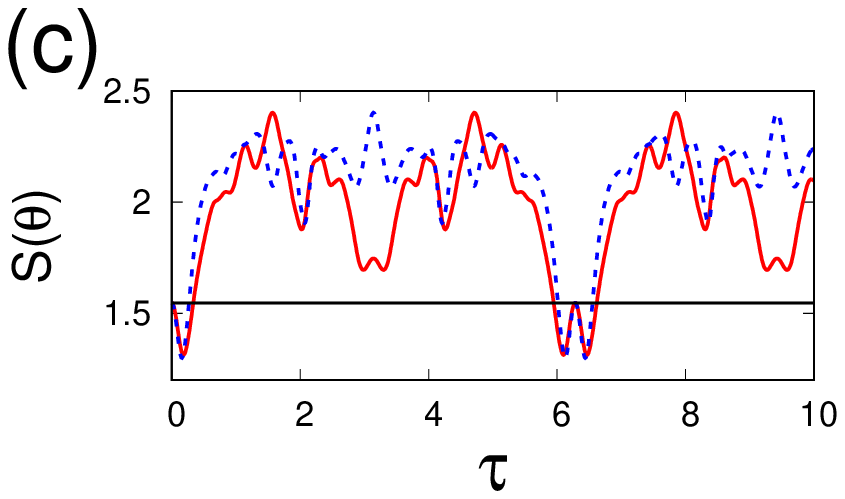}
\includegraphics[scale=0.48]{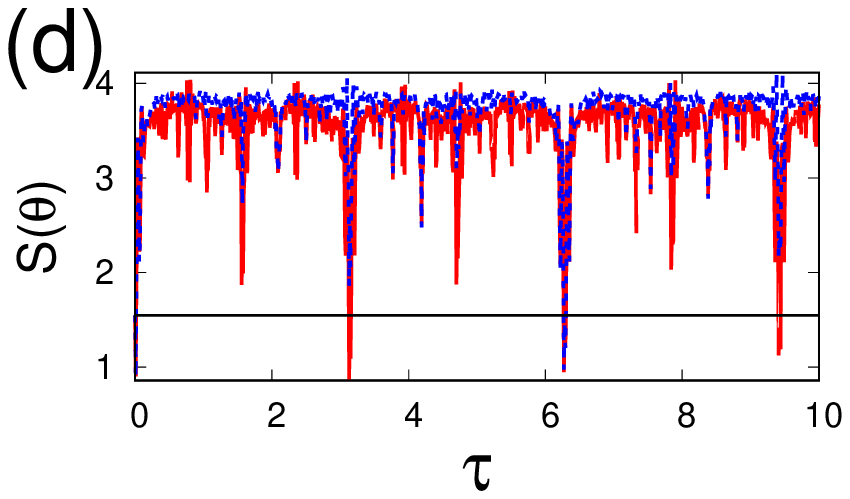}\\ \vspace{-5ex}
\caption{$S(\theta)$ vs $\tau$ for $\Omega = 10^{6}, \,\theta = 0$. $\phi = \tfrac{1}{2} \pi$ (red) and $\phi =  \tfrac{1}{4} \pi$ (blue). $\alpha = 1$ (first column),  $\alpha = 5$  (second column). $\kappa = 0.5$ (top panel), $\kappa = 1$ (bottom panel). The red horizontal line is the reference below which squeezing occurs.}
\label{fig:ent_kappa_param1}
\end{figure}

For $f(N) = (1+\kappa\, a^{\dagger}a)^{1/2}$ ($0\le \kappa \le 1$) and over the same time interval,  there is a remarkable difference in the  qualitative behaviour  in the dynamics of both the tomographic entropy (figures \ref{fig:ent_kappa_param1}(a)-(d)) and the quadrature observables. In particular, squeezing properties are very sensitive to the value of $\kappa$, and even small changes in $\kappa$ lead to substantial changes in them. This is evident in the case of entropic squeezing where for a sufficiently high value of $\alpha$, the extent of squeezing varies significantly with $\kappa$ (figures \ref{fig:ent_kappa_param1}(b), (d) in which  the tomographic entropy is plotted against time for $\Omega = 10^{6}$).  Such sensitive dependence on  $\kappa$ is not observed for $\Omega = 10^{6}/\sqrt{2}$. 

%%%%%%%%%%%%%%%%%%%%%%%%%%%%%%%%%%%%%%%%%%%%%%%%%%%%%%%%%%%%%%%%%%%%%%%%%%
\section{Concluding remarks}
\label{conclusion} 
We have investigated, in some detail, the dynamics of 
the subsystem von Neumann entropy  for the 
different subsystems of a generic optomechanical model. We have established that the  SVNE of the atom collapses to its  {\it maximum}  allowed value over a significant interval of time,  as opposed  to sudden death of entanglement. This novel feature is potentially useful in harnessing entanglement protocols. We have explored experimentally viable parameter regimes and examined how different intensity-dependent 
couplings  affect the nonclassical nature of the radiation field.  We have assessed the extent of nonclassicality of the cavity field 
during time evolution through the  Mandel Q parameter, the Wigner functions and the corresponding optical tomograms. Finally, 
we have quantified entropic and quadrature squeezing properties directly from the field tomograms,  thus by-passing inherently error-prone state reconstruction methods. It is hoped that our study provides  pointers 
to several aspects of the dynamical behaviour of a variety of 
optomechanical systems.

%%%%%%%%%%%%%%%%%%%%%%%%%%%%%%%%%%%%%%%%%%%%%%%%%%%%%%%%%%%%%%%%%%%%%%%%%%%%%%%%%%%%%%%%%%

\appendix
%\section*{Appendices}
\section{The effective Hamiltonian}
\label{appendix_eff_ham}
\renewcommand{\theequation}{A{\arabic{equation}}}
\setcounter{equation}{0}

The Hamiltonian $H$  in \eqref{eqn:parent_hamiltonian} can be written as $H_{0}+H_{1}$, where 
\begin{align}
 H_{0} &=  \omega\, a^{\dagger} a + \omega_{m}\, b^{\dagger} b + 
 \tfrac{1}{2} \omega_{0} \sigma_{z}, \\
 H_{1} &= - G \,  a^{\dagger} a (b + b^{\dagger}) + \Omega\, [a\, f(N) \, \sigma_{+} +  f(N) \,a^{\dagger}\,\sigma_{-}].
 \label{eqn:ham_h0h1}
\end{align}
We express any operator  $A_{I}(t)$ in the interaction picture as 
$e^{i H_{0} t}\, A\, e^{-i H_{0} t}$,  where $A$ is the operator 
in the Schr\"{o}dinger picture, and use  the identities
\begin{equation}
 N^{r}\, a = a(N-1)^{r}, \;\; \sigma_{z}^{r}\, \sigma_{-} = \sigma_{-}\,(\sigma_{z}-2)^{r} \; (r = 1,2,\cdots)
  \label{eqn:nthpower}
\end{equation}
to obtain
\begin{equation}
a_{I} = a\, e^{- i \omega t}, \;\; b_{I} = b\, e^{- i \omega_{m} t}, 
\;\; (\sigma_{-})_{I} = \sigma_{-}\, e^{- i \omega_{0} t}.
\label{eqn:int_absigma}
\end{equation}
Dropping the subscript $I$ and using the resonance condition 
$\omega = \omega_{m} + \omega_{0}$, we get, in terms 
of interaction picture operators, 
\begin{align}
 H_{\textrm{int}} = & \Omega\, \big[a\,f(N)\, \sigma_{+} \,e^{- i \omega_{m} t} + f(N)\, a^{\dagger} \,\sigma_{-}\, e^{ i \omega_{m} t}\big] \nonumber \\[4pt]
                              &\qq\qq - G \,a^{\dagger} a \, \big[b 
                              \,e^{- i \omega_{m} t} + b^{\dagger} \, 
                              e^{ i \omega_{m} t}\big].
\label{eqn:int_hamiltonian}
\end{align}
As is well known, in coarse-grained dynamics in which  
rapidly oscillating terms are neglected in the 
rotating-wave approximation,   a generic Hamiltonian of the form
 \begin{equation}
 \mathcal{H}_{\textrm{I}} = \sum_{n=1}^{N} ( h_{n} \, e^{- i \omega_{n} t} + h_{n}^{\dagger} \, e^{ i \omega_{n} t})
 \label{eqn:gen_int_ham}
 \end{equation}
 can be recast into an  effective Hamiltonian of the form \cite{james}
\begin{equation}
 \mathcal{H}_{\textrm{eff}} = \sum_{n=1}^{N}\sum_{m=1}^{N}  \
(\overline{\omega}_{mn})^{-1}\,[h_{m}^{\dagger}, h_{n}] \, 
e^{i (\omega_{m} - \omega_{n}) t},
 \label{eqn:gen_eff_ham}
 \end{equation}
 where $\overline{\omega}_{mn} = 2\,\omega_{m}\,\omega_{n}/(\omega_{m}+\omega_{n})$.  
We follow a similar procedure, 
 since  $\omega_{m} \gg (G,\, \Omega)$,  and the approximation 
 holds good in this case too.  With the identifications
\begin{equation}
 h_{1} = - G \, a^{\dagger}a\, b,  \;\; h_{2} = \Omega\, a f(N)\, \sigma_{+}, 
 \;\; \overline{\omega}_{mn} = \omega_{m},
\label{eqn:h1h2}
\end{equation}
 it is easy to see that
\begin{align}
 [h_{1}^{\dagger},\, h_{1}] &= G^{2} \,(a^{\dagger}a)^{2},\;\;
  [h_{1}^{\dagger},\, h_{2}] = G \,\Omega \, a\,f(N)\, b^{\dagger},  \\[4pt]
 [h_{2}^{\dagger}, h_{2}] &= -\Omega^{2}\big[a^{\dagger}a f^{2}(n) \, \sigma_{z}  + \big\{a f^{2}(N) a^{\dagger} \nonumber \\
                                       &\hspace{3cm} -  a^{\dagger}a f^{2}(N)\big\} \sigma_{+}  \sigma_{-}\big].
\end{align}
The effective Hamiltonian can now be written as
\begin{align}
 H_{\textrm{eff}}& = \frac{G \Omega}{\omega_{m}} \big[f(N) a^{\dagger}\, b\, \sigma_{-} + a f(N)b^{\dagger} \, \sigma_{+}\big] -  \frac{G^{2}}{\omega_{m}} \,(a^{\dagger}a)^{2}\nonumber \\
  & \hspace{-0.5cm}-  \frac{\Omega^{2}}{\omega_{m}} \big[a^{\dagger}a\, f^{2}(N) \sigma_{z} + \big\{a f^{2}(N)\, a^{\dagger} - a^{\dagger}a\, f^{2}(N)\big\}\, \sigma_{+}\sigma_{-}\big].
\label{eqn:gen_eff_hamiltonian}
\end{align}
To first order in the intensity-dependent coupling $f(N)$ 
(approximating the square $f^{2}(N)$ by unity), we arrive at 
the effective Hamiltonian 
\begin{align}
 H_{\textrm{eff}}  &= \frac{G \Omega}{\omega_{m}} \big[f(N) a^{\dagger}\, b\, \sigma_{-} + a f(N)b^{\dagger} \, \sigma_{+}\big] \nonumber\\[4pt]
 &  - \frac{\Omega^{2}}{\omega_{m}} \big[a^{\dagger} a\, \sigma_{z} - \sigma_{+}\sigma_{-}\big] 
                  -  \frac{G^{2}}{\omega_{m}} \,(a^{\dagger} a)^{2}.
\label{eqn:eff_hamil}
\end{align}
%%%%%%%%%%%%%%%%%%%%%%%%%%%%%%%%%%%%%%%%%%%%%%%%%%%%%%%%%%%%%%%%%%%%%%%%%%
%\section{Appendix B: The state vector}
\section{The state vector}
\label{appendix_state_vector}
\renewcommand{\theequation}{B{\arabic{equation}}}
\setcounter{equation}{0}
The initial state  given by  \eqref{eqn:init_state} is 
\begin{equation}
 \ket{\psi(0)} = \sum_{n=0}^{\infty} l_{n} (\cos\phi \ket{n; 0; e} + \sin\phi \ket{n; 0; g}).
\label{eqn:init_state_vec}
\end{equation}
Its temporal evolution is governed by $H_{\textrm{eff}}$ (\eqref{eqn:eff_hamiltonian}). We have 
\begin{align}
\ket{\psi(t)} = \sum_{n=0}^{\infty} & l_{n} A_{n}(t) \ket{n; 0; e} + \sum_{n=0}^{\infty}  l_{n} B_{n}(t) \ket{n; 0; g} \nonumber \\
                                                             &+ \sum_{n=1}^{\infty} l_{n}C_{n}(t) \ket{n-1; 1; e},
\label{eqn:final_state_vec}
\end{align}
where the time derivatives of the coefficients are obtained from the 
Schr\"{o}dinger equation:  
\begin{align}
\label{eqn:coupled_a}
\dot{A}_{n}(t) &= p A_{n}(t), \\
\label{eqn:coupled_b}
\dot{B}_{n}(t) &= q B_{n}(t) + s C_{n}(t), \\
\label{eqn:coupled_c}
\dot{C}_{n}(t) &= r C_{n}(t) + s B_{n}(t),
%\label{eqn:coupled_eqn}
\end{align}
where
\begin{align}
 p &= i [G \Omega\, n^{2} + \Omega^{2} (n+1)]/{\omega_{m}}, \\
 q &= i [G \Omega\, (n-1)^{2} + \Omega^{2} n]/{\omega_{m}}, \\
 r &= i [G^{2}\, n^{2} - \Omega^{2}\, n]/{\omega_{m}}, \\
 s &= -i G \Omega\, f(n)\sqrt{n}/{\omega_{m}}.
\label{eqn:coefficients}
\end{align}
%with
%\begin{equation}
% A_{n}(0) = \cos\phi, \;\; B_{n}(0) =  \sin\phi, \;\;\textrm{and}\,\, C_{n}(0) = 0,
%\label{eqn:init_conditn}
%\end{equation} 
We impose the initial conditions $A_{n}(0) = \cos\phi$, $B_{n}(0) =  \sin\phi$ and $C_{n}(0) = 0$, and solve the equations given above  to  get
\begin{align}
 \label{eqn:soln_a}
 A_{n}(t) &=  \cos\,\phi\, e^{\,i \gamma_{1} t},\\
 \label{eqn:soln_b}
 B_{n}(t) &=  \sin\,\phi \, \big[\cos\,(R t) + \Delta_{b} \sin\,(R t)\big] 
 \,e^{\,i \gamma_{2} t},\\
 \label{eqn:soln_c}
 C_{n}(t) &=  \sin\phi\,\sin(R t)\, \Delta_{c} \,e^{\,i \gamma_{2} t},
\end{align}
where
\begin{align} \gamma_{1} &= \big[G^{2} n^{2} + (n+1)\Omega^{2}\big]/{\omega_{m}}, \\
 \gamma_{2} &= G^{2}\big(n^{2} - n + \tfrac{1}{2} \big)/{\omega_{m}},\\
 \Delta_{b} &= -i [G^{2}\, (n -\tfrac{1}{2} ) -  \Omega^{2}\, n]/{(R\omega_{m})}, \\
 \Delta_{c} &= -i G\,\Omega\,\sqrt{n}\, f(n)/{(R\omega_{m})},\\
 R  &= \Big\{\tfrac{1}{4}G^{4}(2n^{2} - 2n+1)^{2} + G^2\Omega^{2}\, n\, f^{2}(n)  \nonumber\\
     &\q - [G^{2}\, (n-1)^{2} +  \Omega^{2}\, n] [G^{2} n^{2} -  \Omega^{2}\, n] \Big\}^{1/2}/{\omega_{m}}. 
 \label{eqn:parameters}
\end{align}
The respective reduced density matrices for the atom, the mirror and the field  are then given by   
\begin{align}
\label{eqn:red_den_mat_a}
 \rho_{a} &= \sum_{n=0}^{\infty} \Big[\Big(| l_{n}|^{2} \, |A_{n}|^{2} + | l_{n+1}|^{2} |C_{n+1}|^{2} \Big) \ket{e}\bra{e} \nonumber \\
 & + | l_{n}|^{2} \Big(
  |B_{n}|^{2} \ket{g}\bra{g} 
 +  A_{n}  B_{n}^{*} \ket{e}\bra{g} + A_{n}^{*} B_{n} \ket{g}\bra{e} \Big)\Big], 
\end{align}
\begin{align}
\label{eqn:red_den_mat_m}
 \rho_{m} &=  \sum_{n=0}^{\infty} \Big[ | l_{n}|^{2} \Big( |A_{n}|^{2} + |B_{n}|^{2} \Big)\ket{0}\bra{0} + | l_{n+1}|^{2} |C_{n+1}|^{2} \ket{1}\bra{1} \nonumber \\
 &\q\,\,\,\,+  l_{n} l_{n+1}^{*} A_{n}C_{n+1}^{*} \ket{0}\bra{1} + l_{n}^{*}  l_{n+1}A_{n}^{*}  C_{n+1} \ket{1}\bra{0} \Big],
\end{align}
\begin{align}
\label{eqn:red_den_mat_f}
 \rho_{f} &=   \sum_{n=0}^{\infty}\sum_{m=0}^{\infty}  l_{n} l_{m}^{*} [ A_{n}A_{m}^{*}  + B_{n}B_{m}^{*} ] \ket{n}\bra{m}  \nonumber \\
 &\qq+ \sum_{n=1}^{\infty}\sum_{m=1}^{\infty} l_{n} l_{m}^{*} C_{n}C_{m}^{*} \ket{n-1}\bra{m-1},
%\label{eqn:red_den_mat_a}
\end{align}
where the coefficients $A_{n}, \,
B_{n}, \,C_{n}$ have the  time dependences indicated  
in \eqref{eqn:soln_a}--\eqref{eqn:soln_c}. 

%%%%%%%%%%%%%%%%%%%%%%%%%%%%%%%%%%%%%%%%%%%%%%%%%%%%%%%%%%%%%%%%%%%%%%%%%%
%\section*{Appendix C: The Wigner function for the field}

\section{The Wigner function for the field}
\label{appendix_wigner_function}
\renewcommand{\theequation}{C{\arabic{equation}}}
\setcounter{equation}{0}

The Wigner function $W(p, \,q)$ for a state with density matrix $\rho_{f}$ is defined by the integral \cite{safari}
\begin{equation}
W(p, \,q) = \frac{1}{\pi} \int_{-\infty}^{\infty} \, dq' \expect{q+q'}{\rho_{f}}{q-q'} e^{-2i \, p\, q'}
\label{eqn:wigner_phase_space}
\end{equation}
in standard notation.
We now use the fact that $e^{-i\,\hat{p}\, x} \ket{q} = \ket{q+x}$, and the identity $ \displaystyle{e^{\hat{A}+\hat{B}} = e^{-\frac{1}{2} [\hat{A},\, \hat{B}]}\, e^{\hat{A}} \,e^{\hat{B}}}$ (since 
$[\hat{A},\,\hat{B}]$ is proportional to the unit operator 
in the cases of interest here).   It is straightforward to see that 
\begin{align}
D(\alpha) \ket{-q'} &= e^{-\frac{1}{2}i\,qp} e^{i\,p\,\hat{q}} e^{- i\,q\,\hat{p}} \ket{q - q'}\nonumber \\
                             &= e^{\frac{1}{2}i\,qp} e^{- i\,p\,q'} \ket{q - q'},
\label{eqn:wigner_d_ket_q}
\end{align}
where  $D(\alpha) = e^{\alpha a^{\dagger} - \alpha^{*} a} = 
e^{ip\hat{q} - iq\hat{p}}$ is the displacement operator. 
Inverting \eqref{eqn:wigner_d_ket_q},  we have
\begin{equation}
 \ket{q - q'} = e^{-i\,q\,p/2}\, e^{i\,p\,q'} \, D(\alpha) \ket{-q'}.
\label{eqn:wigner_d_ket_q-x}
\end{equation}
Substituting \eqref{eqn:wigner_d_ket_q-x} and its Hermitian conjugate  in \eqref{eqn:wigner_phase_space}, we get
\begin{align}
 W(\alpha) &=  (1/\pi) \int_{-\infty}^{\infty} \, dq' \expect{q'}{D^{\dagger}(\alpha) \, \rho_{f}\, D(\alpha)  }{-q'} \nonumber\\
                  &= (2/\pi) \int_{0}^{\infty} \, dq' \expect{q'}{D^{\dagger}(\alpha) \, \rho_{f}\, D(\alpha)  }{-q'}.
\label{eqn:wig_d_alpha_step1}
\end{align}
Inserting the identity operator $\sum_{n=0}^{\infty} \ket{n}\bra{n}$ in \eqref{eqn:wig_d_alpha_step1} and using the fact that $\inner{n}{-q} = (-1)^{n}\inner{n}{q}$, we obtain
\begin{equation}
 W(\alpha) = (2/\pi) \sum_{n=0}^{\infty} (-1)^{n} \expect{n; \,\alpha}{\rho_{f}}{n; \,\alpha},
\label{eqn:wigner_def_re}
\end{equation}
where $\ket{n;\, \alpha} = D(\alpha)\ket{n}$ is a displaced number state 
(or a generalized coherent state). Using \eqref{eqn:wigner_def_re},  the expression 
in \eqref{eqn:wigner_cav}
for the Wigner function $W_{f}(\alpha)$ 
corresponding to the density matrix $\rho_{f}$  
given by \eqref{eqn:red_den_mat_f} is obtained easily.   
%\begin{align}
%  W_{f}(\alpha) = \frac{2}{\pi} &\sum_{n,k,l=0}^{\infty}(-1)^{n}\, q_{k}\, q_{l}^{*} \bigg\{ [ A_{k}\, A_{l}^{*}  + B_{k}\, B_{l}^{*}] \expect{n}{D^{\dagger}}{k} \expect{l}{D^\dagger}{n} \nonumber \\
%  &\q+ C_{k}\, C_{l}^{*} \expect{n}{D^{\dagger}}{k-1} \expect{l-1}{D^{\dagger}}{n}\bigg\},
%\label{eqn:wigner_cav}
%\end{align}
%with the identification that
%\begin{equation}
% \expect{n}{D}{m} = 
%  \left\{
%    \begin{array}{l}
%      e^{-\frac{1}{2}|\alpha|^{2}}\, \sqrt{\frac{m!}{n!}}\,\, \alpha^{n-m} \, L_{m}^{n-m}(|\alpha|^{2}) \qquad\qquad \textrm{for}\; \;  n\ge m,\\
%      e^{-\frac{1}{2}|\alpha|^{2}}\, \sqrt{\frac{n!}{m!}}\, \,(-\alpha^{*})^{m-n} \, L_{n}^{m-n}(|\alpha|^{2}) \qquad\; \textrm{for}\;\; n<m,
%    \end{array}
%  \right.
%\end{equation}
%and $L_{n}^{m}(x)$ is the associated Laguerre polynomial.
%%%%%%%%%%%%%%%%%%%%%%%%%%%%%%%%%%%%%%%%%%%%%%%%%%%%%%%%%%%%%%%%%%%%%%%%%%%%%%%%%%%%%%%%%%%%%%

% Bibliography
\bibliography{reference}
%
% ****** End of file apssamp.tex ******

\end{document}